\documentclass[
    floats,
    floatfix,
    amssymb,
    prd,
    twocolumn,
    superscriptaddress,
    nofootinbib,
    unicode,
    colorlinks=true,
    linktocpage=true
]{revtex4-2}
\PassOptionsToPackage{dvipsnames,table}{xcolor}

\usepackage{graphicx, amssymb, amsmath, amsfonts, epsfig, fixmath}
\usepackage{epstopdf}
\usepackage[normalem]{ulem}
\usepackage{tabularx}
\usepackage{longtable}
\usepackage{bm}
\usepackage{dcolumn}
\usepackage{rotating}
\usepackage{longtable}
\usepackage{comment}
\usepackage{enumerate}
\usepackage[unicode]{hyperref}
\usepackage{orcidlink}
\usepackage{slashed}
\usepackage{tensor, multirow}
\usepackage{url}
\usepackage{xcolor}
\usepackage[caption=false]{subfig}
\usepackage{float}
\usepackage{lipsum}
\usepackage{cancel}

\hypersetup{
    colorlinks=true, 
    citecolor=MidnightBlue,
    linkcolor=MidnightBlue, 
    urlcolor=MidnightBlue, 
    linktocpage=true
}

\newcommand{\sissa}{SISSA, Via Bonomea 265, 34136 Trieste, Italy \&  INFN Sezione di Trieste}

\newcommand{\ifpu}{IFPU - Institute for Fundamental Physics of the Universe, Via Beirut 2, 34014 Trieste, Italy}

\newcommand{\inaf}{INAF - Osservatorio Astronomico di Cagliari, Via della Scienza 5, 09047 Selargius, Italy}

\newcommand{\mpifr}{Max-Planck-Institut f\"{u}r Radioastronomie, Auf dem H\"{u}gel 69, D-53121 Bonn, Germany}
\newcommand{\aei}{Max-Planck-Institut f\"{u}r Gravitationsphysik (Albert-Einstein-Institut), Leibniz Universit\"{a}t Hannover, Callinstra{\ss}e 38, D-30167 Hannover, Germany}

\newcommand{\iisertvm}{School of Physics, Indian Institute of Science Education and Research Thiruvananthapuram, Maruthamala P.O., Thiruvananthapuram, Kerala 695551, India}
\newcommand{\bupi}{Fakul\"at f\"ur Physik, Universit\"at Bielefeld, Postfach 100131, 33501 Bielefeld, Germany}

\begin{document}

\title{Constraints on Einstein-\ae ther gravity from the precision timing of PSR J1738+0333}

\author{Massimo~Vaglio~\orcidlink{0000-0002-7285-3489}}
\affiliation{\sissa}
\affiliation{\ifpu}

\author{Amodio~Carleo~\orcidlink{0000-0001-9929-2370}}
\affiliation{\inaf}

\author{Abhimanyu~Susobhanan~\orcidlink{0000-0002-2820-0931}}
\affiliation{\iisertvm}
\affiliation{\aei}

\author{Enrico~Barausse~\orcidlink{0000-0001-6499-6263}}
\affiliation{\sissa}
\affiliation{\ifpu}

\author{Bilel Ben Salem~\orcidlink{0000-0001-7131-4929}}
\affiliation{\bupi}

\author{Alessandro~Corongiu~\orcidlink{0000-0002-5924-3141}}
\affiliation{\inaf}

\author{Paulo~C.~C.~Freire~\orcidlink{0000-0003-1307-9435}}
\affiliation{\mpifr}

\author{Delphine~Perrodin~\orcidlink{0000-0002-8509-5947}}
\affiliation{\inaf}

\begin{abstract}
We constrain Einstein-\ae ther gravity -- a Lorentz-violating extension of General Relativity in which a dynamical, unit timelike vector field selects a preferred frame -- using updated high-precision pulsar timing observations of PSR J1738+0333 from EPTA second Data Release and the NANOGrav 9-year release, in combination with ToAs from Arecibo, Green Bank, Nançay, Parkes, and Westerbork. 
Our method accounts for both conservative and dissipative first post-Newtonian corrections arising from Lorentz violation; here we apply it to PSR J1738+0333 using the Bayesian timing pipeline Vela to process the full ToA dataset. We sample the joint posterior over binary component masses, post-Keplerian parameters and center-of-mass velocity components, and then apply a resampling scheme to propagate posteriors into robust constraints on the fundamental theory parameters, obtaining the most stringent strong-field bounds on the Einstein-æther coupling constants from a single binary pulsar system to date.

\end{abstract}

\maketitle

\section{Introduction}
The discovery of pulsars has marked a turning point in the field of radio astronomy and gravitational physics~\cite{Hewish:1968bj}. Only a few years after this landmark breakthrough, the observation of the first pulsar within a binary system~\cite{Hulse:1974eb}, enabled the first precision tests of the radiative sector of General Relativity (GR), probing a regime far more relativistic than solar system or laboratory experiments could access~\cite{Taylor:1982zz}. Today, over one hundred such compact binary systems are known~\cite{Manchester:2004bp}. 
With the advent of precision tests of gravity, the fundamental pillars of GR have been subjected to increasingly stringent scrutiny~\cite{Will:2014kxa, Berti:2015itd}. Among these pillars is local Lorentz invariance — equivalently, the absence of preferred reference frames — for gravitational phenomena. Despite this progress, while violations of Lorentz invariance are extremely tightly constrained in the matter sector~\cite{Kostelecky:2003fs, Kostelecky:2008ts}, the Lorentz symmetry of the gravitational interaction is subject to comparatively weaker bounds~\cite{Mattingly:2005re, Will:2014kxa}.

Einstein-\ae ther gravity was proposed in~\cite{Jacobson:2000xp} to provide a fully covariant framework in which Lorentz violations are implemented through a dynamical unit timelike vector field, allowing one to analyze their degrees of freedom and phenomenology using standard gravitational tools. Interest in this theory is further motivated by ultraviolet considerations: allowing time and space to scale differently at high energies provides a possible route toward improving the renormalization properties of gravity, at the price of breaking Lorentz invariance~\cite{Anselmi:2007ri, Horava:2009uw}. The most prominent realization of this idea is Ho\v{r}ava gravity~\cite{Horava:2009uw}; when the \ae ther is restricted to being hypersurface-orthogonal, Einstein--\ae ther theory reproduces its infrared limit (khronometric theory~\cite{Blas:2010hb}), reinstating general covariance while capturing its low-energy dynamics.

In Einstein-\ae ther theory, Lorentz violations can lead to several observable consequences. At the level of gravitational wave propagation, the theory predicts tensor, vector, and scalar modes traveling at generally different speeds; the tensor mode speed is however tightly constrained by the multimessenger observation of GW170817~\cite{Monitor:2017mdv}. A distinct and complementary class of effects arises in the dynamics of strongly gravitating systems. Although the theory preserves the weak equivalence principle — so that test bodies follow universal trajectories — the coupling between the \ae ther field and the metric causes compact objects to respond differently to gravity depending on their gravitational binding energy~\cite{Foster:2007gr,Yagi:2013qpa,Yagi_2014,Barausse:2019yuk,Gupta_2021}, violating the strong equivalence principle (SEP) (as in scalar--tensor theories) and making relativistic binary pulsars particularly sensitive laboratories for testing Einstein--\ae ther gravity.

Pulsars in binary and triple systems serve as high-precision cosmic clocks that allow for some of the most stringent tests of GR and constrain possible deviations from it~\cite{Carleo:kiselev,Carleo:non_local,Freire:2024, Freire:2024, ben2023tests}. The large orbital separation (compared to their size) allows one to treat both bodies as point-like masses, neglecting contributions like tidal interactions, yet their compactness keeps the system sensitive to strong-field gravitational effects. Among these, double neutron star (NS) systems stand out as especially powerful probes of relativistic gravity, with systems such as the Double Pulsar providing some of the most stringent tests of GR~\cite{Kramer:2004hd, Hu_2022}. Pulsars with white dwarf (WD) companions are, however, particularly valuable for testing alternative theories of gravity: their strong asymmetry in gravitational self-energy enhances predicted beyond-GR effects, most notably dipole gravitational radiation, which vanishes for equal-mass binaries. They currently provide the tightest constraints on the parameter space of alternative theories of gravity~\cite{Freire:2024,Hu:2025nua}, and their constraining power could further improve with the Square Kilometer Array (SKA)~\cite{SKA_Vivek}. 

In the specific case of Einstein-\ae ther theory, previous analyses have been incomplete in complementary ways. Some works considered only preferred-frame corrections to conservative orbital dynamics, neglecting dissipative effects entirely~\cite{Bell_1996,Shao_2013}. Others focused on a single beyond-GR effect — most often the drift of the binary orbital period — while ignoring other corrections to the binary dynamics~\cite{Yagi:2013qpa,Yagi_2014,Gupta_2021}. Moreover, the orbital period derivative was typically treated as an independent observable, rather than being modeled as a function of the masses and orbital parameters and varied self-consistently. Although often acceptable as a first approximation, this approach neglects parameter correlations and discards information carried by additional timing effects, yielding a suboptimal analysis.

In this paper, we present a comprehensive Bayesian framework for analyzing pulsar timing data within Einstein--\ae ther gravity, accounting simultaneously for both conservative and dissipative beyond-GR effects at first post-Newtonian order (1PN), including preferred-frame contributions to the orbital dynamics and corrections to pulse propagation. We apply this framework to PSR J1738+0333~\cite{Freire:2007sb}, currently one of the most precisely timed pulsar-WD system, for which the orbital period derivative is the dominant measurable beyond-GR effect. By sampling the joint posterior over binary masses, post-Keplerian parameters, Einstein--\ae ther coupling constants, and center-of-mass velocity components, we properly account for parameter correlations and obtain more robust bounds on the theory's parameter space. 

In Section~\ref{sec:Einstein-Aether} we introduce Einstein--\ae ther gravity and its action. In Section~\ref{sec:two_body} we summarize the two-body dynamics, deriving the conservative 1PN equations of motion, the secular evolution of the orbital elements, and the explicit expressions for the post-Keplerian parameters relevant to our analysis. Section~\ref{sec:obs_and_dataset} describes the dataset and the Bayesian timing pipeline used to process the PSR J1738+0333 observations. Section~\ref{sec:results} presents our results and constraints on the Einstein--\ae ther coupling constants. 
We summarize our conclusions in Section~\ref{sec:conclusions}.
We use the $(+---)$ signature throughout.

\section{Einstein-\AE ther gravity}
\label{sec:Einstein-Aether}
Einstein--\ae ther gravity was first introduced by Jacobson and Mattingly
in~\cite{Jacobson:2000xp}. Its action is given by
\begin{align}\label{eq:action}
S_{\AE} = -\frac{1}{16\pi G} \int \Big(
R + \frac{1}{3} c_\theta \theta^2
+ c_\sigma \sigma_{\mu\nu}\sigma^{\mu\nu}
+ c_\omega \omega_{\mu\nu}\omega^{\mu\nu} \nonumber \\
\Big. + c_a A_\mu A^\mu
+ \lambda (U_\mu U^\mu - 1)
\Big)\sqrt{-g}\, d^4x
+ S_{\text{mat}}(\psi, g_{\mu\nu}) .
\end{align}
The terms in the Lagrangian involving derivatives of the \ae ther field, $\nabla_\mu U_\nu$,  are decomposed into the expansion \( \theta \), shear \( \sigma_{\mu\nu} \), vorticity
\( \omega_{\mu\nu} \), and acceleration \( A_\mu \), defined as
\begin{align}
\theta &= \nabla_\mu U^\mu , \quad
A_\mu = U^\nu \nabla_\nu U_\mu , \\
\sigma_{\mu\nu} &= \nabla_{(\nu} U_{\mu)} + A_{(\mu} U_{\nu)}
- \frac{1}{3}\theta\, h_{\mu\nu} , \\
\omega_{\mu\nu} &= \nabla_{[\nu} U_{\mu]} + A_{[\mu} U_{\nu]} ,
\end{align}
where \( h_{\mu\nu} = g_{\mu\nu} - U_\mu U_\nu \) denotes the induced metric,
acting as a projector onto the hypersurface orthogonal to \( U^\mu \).

By varying the action~\eqref{eq:action} with respect to the metric
$g_{\mu\nu}$, the \ae ther field $U^\mu$, and the Lagrange multiplier
$\lambda$, and eliminating the latter from the resulting equations,
one obtains the generalized Einstein equations
\begin{equation}\label{eq:Eeq}
E_{\alpha\beta} \equiv G_{\alpha\beta} - T^{\AE}_{\alpha\beta}
- 8\pi G\, T^{\rm mat}_{\alpha\beta} = 0\,,
\end{equation}
and the \ae ther equations
\begin{equation}\label{eq:AEeq}
\AE_\mu \equiv \left[
\nabla_\alpha J^{\alpha\nu}
- \left(c_a - \frac{c_\sigma + c_\omega}{2}\right)
A_\alpha \nabla^\nu U^\alpha
\right] h_{\mu\nu} = 0\,,
\end{equation}
where $G_{\alpha\beta}$ is the Einstein tensor, and the \ae ther
stress--energy tensor is
\begin{align}\label{eq:Tae}
T^{\AE}_{\alpha\beta} &=
\nabla_\mu\!\left(
J_{(\alpha}{}^\mu U_{\beta)}
- J^\mu{}_{(\alpha} U_{\beta)}
- J_{(\alpha\beta)} U^\mu
\right) \nonumber \\
&\quad + \frac{c_\omega + c_\sigma}{2}
\left[
(\nabla_\mu U_\alpha)(\nabla^\mu U_\beta)
- (\nabla_\alpha U_\mu)(\nabla_\beta U^\mu)
\right] \nonumber \\
&\quad + U_\nu(\nabla_\mu J^{\mu\nu})\, U_\alpha U_\beta\nonumber\\ &\quad 
- \left(c_a - \frac{c_\sigma + c_\omega}{2}\right)
\!\left(A^2 U_\alpha U_\beta - A_\alpha A_\beta\right) \nonumber \\
&\quad + \frac{1}{2}
M^{\sigma\rho}{}_{\mu\nu}
\nabla_\sigma U^\mu \nabla_\rho U^\nu\, g_{\alpha\beta}\,,
\end{align}
with the definitions
\begin{align}
J^\alpha{}_\mu &\equiv M^{\alpha\beta}{}_{\mu\nu}\nabla_\beta U^\nu\,, \\
M^{\alpha\beta}{}_{\mu\nu} &\equiv
\left(\frac{c_\sigma + c_\omega}{2}\right) h^{\alpha\beta} g_{\mu\nu}
+ \left(\frac{c_\theta - c_\sigma}{3}\right) \delta^\alpha_\mu \delta^\beta_\nu
\nonumber\\
&\quad
+ \left(\frac{c_\sigma - c_\omega}{2}\right) \delta^\alpha_\nu \delta^\beta_\mu
+ c_a\, U^\alpha U^\beta g_{\mu\nu}\,.
\end{align}
Finally, the matter stress--energy tensor is defined in the standard way,
\begin{equation}\label{eq:Tmat}
T^{\alpha\beta}_{\rm mat} \equiv
-\frac{2}{\sqrt{-g}}\frac{\delta S_{\rm mat}}{\delta g_{\alpha\beta}}\,.
\end{equation}

This theory possesses propagating degrees of freedom corresponding to a
transverse--traceless tensor mode, a transverse vector mode, and a scalar
mode~\cite{Jacobson:2004ts, Dong:2026xab}. The transverse--traceless mode corresponds
to the propagation of gravitational waves with speed
\begin{equation}\label{eq:cT}
c_T^2 = \frac{1}{1 - c_\sigma}\,.
\end{equation}
The remaining propagation speeds are obtained similarly by linearizing
the field equations and read
\begin{align}
c_V^2 &=
\frac{c_\sigma + c_\omega - c_\sigma c_\omega}
{2 c_a (1 - c_\sigma)}
\,, \label{eq:cV}\\[6pt]
c_S^2 &=
\frac{(c_\theta + 2 c_\sigma)\left(1 - \tfrac{1}{2} c_a\right)}
{3 c_a (1 - c_\sigma)\left(1 + \tfrac{1}{2} c_\theta\right)}
\,, \label{eq:cS}
\end{align}
where the subscripts $V$ and $S$ denote vector and scalar, respectively.

A number of experimental and theoretical results constrain these
parameters. Requiring the absence of gradient instabilities and ghosts
imposes $c_T^2 > 0$, $c_V^2 > 0$, and
$c_S^2 > 0$~\cite{Jacobson:2004ts,Garfinkle:2011iw}, while requiring
the modes to carry positive energy gives $c_a > 0$ and
$c_\omega > 0$~\cite{eling:2005zq}.
Furthermore,  subluminal graviton propagation would cause
ultrarelativistic matter to lose energy via a Cherenkov-like process;
since this is not observed in ultrahigh-energy cosmic rays, one requires
$c_I^2 \gtrsim 1 - \mathcal{O}(10^{-15})$ for
$I = T, V, S$~\cite{Elliott:2005va}.
The coincident detection of GW170817 and GRB170817A constrained the tensor speed to
$-3\times10^{-15} < c_T - 1 < 7\times10^{-16}$~\cite{Monitor:2017mdv},
implying $c_\sigma \sim \mathcal{O}(10^{-15}) \simeq 0$, so that the
theory is effectively characterized by three independent parameters.

To facilitate comparison with the Parametrized Post-Newtonian (PPN)
formalism~\cite{Will:2014kxa,muller:2005sr}, it is convenient to work
with the parameter set $\{\alpha_1, \alpha_2, c_\omega\}$, where
\begin{align}
\alpha_1 &=
\frac{4 c_\omega (c_a - 2 c_\sigma) + c_a c_\sigma}
{c_\omega (c_\sigma - 1) - c_\sigma}
\;\simeq\; -4 c_a\,, \label{eq:alpha1_def}\\[6pt]
\alpha_2 &=
\frac{\alpha_1}{2}
+ \frac{3 (c_a - 2 c_\sigma)(c_\theta + c_a)}
{(2 - c_a)(c_\theta + 2 c_\sigma)}
\;\simeq\;
\frac{\alpha_1}{2}
+ \frac{3 c_a\left(1 + \tfrac{c_a}{c_\theta}\right)}{2 - c_a}\,.
\label{eq:alpha2_def}
\end{align}
The parameter $\alpha_1$ characterizes preferred-frame effects
that depend on the magnitude of the system's velocity relative to the
preferred frame, while $\alpha_2$ controls effects that depend on the
orientation of this velocity~\cite{1972ApJ...177..757W, Foster:2005dk}. Solar system tests
require $|\alpha_1| \lesssim 10^{-4}$ and
$|\alpha_2| \lesssim 10^{-7}$~\cite{Will:2014kxa,muller:2005sr}.

\section{Two-Body Dynamics in Einstein–\ae ther and Post-Keplerian Parameters}

\label{sec:two_body}

\subsection{Strong-equivalence principle violation and sensitivities}
In Einstein-\ae ther theory, matter is assumed to couple only to the metric tensor, with no direct coupling to the \ae ther. This confines Lorentz violations to the gravitational sector, leaving particle physics experiments unaffected. Nevertheless, strongly self-gravitating compact objects such as NSs have a non-negligible fraction of their mass in the form of gravitational binding energy. As a result, their mass is influenced by the presence of the \ae ther field, leading to an effective coupling (``sensitivity'') between compact objects and the \ae ther through their self-energy~\cite{Eardley:1975fgi, Damour:1996ke}.

\begin{figure*}[t] 
    \centering
    \includegraphics[width=0.99\textwidth]{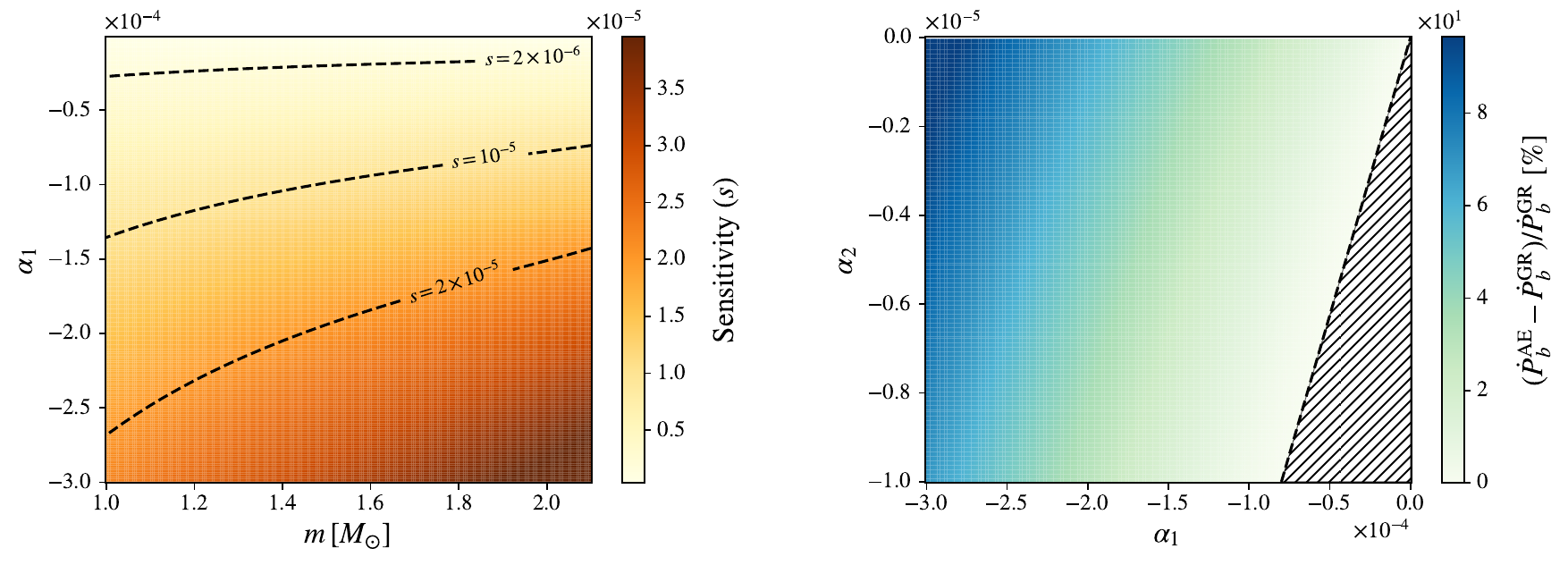}
    \caption{Sensitivity parameter $s$ of a NS as a function of mass for fixed $\alpha_2 = 0$ and $c_\omega = 10^{-3}$ (left panel), and relative deviation of the predicted orbital period derivative in Einstein-\ae ther theory from the GR prediction for PSR J1738$+$0333 as a function of $\alpha_1$ and $\alpha_2$ for $c_\omega = 10^{-3}$ and $m_p=1.42M_\odot$(right panel). In both panels the NS sensitivity is computed assuming the APR~\cite{Akmal:1998cf} equation of state, while zero sensitivity is assumed for the companion. The dashed contours in the left panel mark constant values of $s$. In the right panel, the dashed line indicates the theoretical boundary $\alpha_2 = \alpha_1/8$, below which the theory is not defined (see Sec.~\ref{sec:post-Keplerian}); the hatched region is therefore excluded.}
    \label{fig:sensitivity_and_pb_dot}
\end{figure*}

Following~\cite{Foster:2007gr}, the standard geodesic point-particle action is 
generalized to account for the sensitivities as follows:
\begin{equation}
\label{eq:actionpp}
S_{{\rm pp}\,,A} = - \int d\tau_A \; \tilde{m}_{A}(\gamma_{A})\,,
\end{equation}
where $A$ labels the object, $\tilde{m}_A$ is the mass of the object, $d\tau_A$ is the proper time along the 
$A$-th worldline and $\gamma_{A} \equiv U_{\mu} u_{A}^{\mu}$
is the Lorentz 
factor of the body with respect to the \ae ther, defined through the contraction 
of the \ae ther field $U^{\mu}$ with the particle four-velocity $u_{A}^{\mu}$.
Since the PN expansion models the dynamics as an expansion in the orbital velocity, 
and $\gamma_A \approx 1$ for bodies moving slowly relative to the \ae ther, 
Eq.~\eqref{eq:actionpp} can be expanded as
\begin{align}
\label{taylor-pp}
S_{{\rm pp}\,A} &= -\tilde{m}_{A} \int d\tau_A 
\left\{ 1 + \sigma_{A}(1 - \gamma_{A}) \right.
\nonumber \\
&\left. + \frac{1}{2}\sigma_{A}'(1 - \gamma_{A})^{2} 
+ \mathcal{O}\left[(1 - \gamma_{A})^{3}\right] \right\}\,,
\end{align}
where $\tilde{m}_{A} \equiv \tilde{m}_{A}(1)$ is the rest mass and the two 
sensitivity parameters $\sigma_{A}$ and $\sigma_{A}'$ are defined as
\begin{align}
\label{sigma-def}
\sigma_{A} &\equiv -\left.\frac{d\ln\tilde{m}_{A}(\gamma_A)}{d\ln\gamma_{A}}
\right|_{\gamma_{A}=1}\,,\\[4pt]
\sigma_{A}' &\equiv \sigma_{A} + \sigma_{A}^{2} + 
\left.\frac{d^{2}\ln\tilde{m}_{A}(\gamma_A)}{d(\ln\gamma_{A})^{2}}
\right|_{\gamma_{A}=1}\,.
\end{align}
A rescaled version of the leading sensitivity is also useful, 
\begin{equation}
\label{s-sigma-rel}
s_{A} \equiv \frac{\sigma_{A}}{1 + \sigma_{A}}\,.
\end{equation}
All these quantities vanish for bodies with negligible gravitational 
binding energy, recovering the geodesic limit. For NSs, however, 
$s_A (\sigma_A)$ is generically non-zero, reflecting  violations of the  SEP. Finally, the effective \ae ther--matter coupling induced by Eq.~\eqref{eq:actionpp} 
modifies the \ae ther field equations~\eqref{eq:AEeq} by introducing a 
worldline source proportional to the sensitivities,
\begin{multline}
\label{eq:AEsens}
\tilde{\AE}^{\mu} \equiv \AE^{\mu} + \frac{8\pi G\tilde{m}_A}{u^{0}\sqrt{-g}}
\times \\ 
\delta^{(3)}(x^{i} - x^{i}_{A})\left(\sigma_A + \sigma_A'(1-\gamma_A)\right)
\left(u_{A}^{\mu} - \gamma_A U^{\mu}\right) = 0\,,
\end{multline}
where $x_{A}^{i}$ is the worldline of the $A$-th body.

\subsection{Conservative 1PN Dynamics}

In this section, we summarize the conservative post-Newtonian  dynamics of a compact binary in Einstein--\ae ther  theory, following~\cite{Gupta_2021} and~\cite{Will_2018}. We work in the preferred frame in which the \ae ther is asymptotically at rest and retain terms through the first post-Newtonian (1PN) order. We introduce the Newtonian gravitational constant $G_N$, as measured through Cavendish-type
experiments~\cite{Carroll:2004ai, Yagi_2014}, which is related to the bare constant
 $G$ appearing in Eq.~\eqref{eq:action} by
\begin{equation}
G_N=\frac{G}{1-\frac{c_a}{2}}
\label{eq:Cavendish}
\end{equation}

First, we consider two compact objects labeled by $A=\{1,2\}$, with coordinate positions $\mathbf{x}_A$, velocities $\mathbf{v}_A=\dot{\mathbf{x}}_A$, and bare masses $\tilde m_A$, and we define
\begin{align}
&r_A = |\mathbf{x}-\mathbf{x}_A|, \quad
\mathbf{n}_A=\frac{\mathbf{x}-\mathbf{x}_A}{r_A},\quad
\mathbf n_{AB}=\frac{\mathbf x_A-\mathbf x_B}{r_{AB}}, \nonumber \\
&r=|\mathbf{x}_1-\mathbf{x}_2|, \quad
\mathbf{n}=\frac{\mathbf{x}_1-\mathbf{x}_2}{r}, \quad
\mathbf{v}_{21}=\mathbf{v}_2-\mathbf{v}_1,
\end{align}
where $\mathbf{x}$ represents the field point. To 1PN order, the metric generated by the two bodies reads
\begin{align}
&g_{00} =\; 1
-\frac{2G_N\tilde m_1}{c^2 r_1} +\frac{1}{c^4}\Bigg[
\frac{2G_N^2\tilde m_1^2}{r_1^2} +\frac{2G_N^2\tilde m_1\tilde m_2}{r_1 r_2} \nonumber\\
&
+\frac{2G_N^2\tilde m_1\tilde m_2}{r_1 r}
-\frac{3G_N\tilde m_1}{r_1}v_1^2(1+\sigma_1)
\Bigg]
+1\leftrightarrow2 ,
\label{eq:g00_1PN}
\end{align}
\begin{align}
g_{0i}
=
&-\frac{1}{c^3}\left[
B_1^-\frac{G_N\tilde m_1}{r_1}v_1^i\right.\nonumber\\
&\left.+B_1^+\frac{G_N\tilde m_1}{r_1}
(\mathbf v_1\!\cdot\!\mathbf n_1)n_1^i
\right]
+1\leftrightarrow2 ,
\label{eq:g0i_1PN}
\end{align}
\begin{equation}
g_{ij}
=
-\left(1+\frac{2G_N\tilde m_1}{c^2 r_1}\right)\delta_{ij}
+1\leftrightarrow2 .
\label{eq:gij_1PN}
\end{equation}
The \ae ther field, instead, is given by
\begin{equation}
U^0 = 1 + \frac{G_N\tilde m_1}{c^2 r_1} + 1\leftrightarrow2,
\end{equation}
\begin{equation}
U^i
=\frac{1}{c^3}\frac{G_N\tilde m_1}{r_1}
\left(
C_1^- v_1^i
+ C_1^+ (\mathbf v_1\!\cdot\!\mathbf n_1)n_1^i
\right)
+1\leftrightarrow2 .
\end{equation}
where the coefficients $B_A^\pm$ and $C_A^\pm$ depend on the \ae ther coupling constants, $\alpha_1,\alpha_2$ and $c_\omega$ and on the sensitivities $\sigma_A$. 

It is convenient to introduce the active gravitational mass
\begin{equation}\label{eq:active_mass}
m_A=\tilde m_A(1+\sigma_A).
\end{equation}
which makes apparent how the sensitivity parameterizes violations of the SEP.
Inserting the PN-expanded fields in the equations of motion Eqs.~\eqref{eq:Eeq}-\eqref{eq:AEeq}, one finds that at leading (Newtonian) order
\begin{equation}\label{eq:newtonian}
\frac{d\mathbf v_A}{dt}
=
-\frac{\mathcal{G} m_B}{r^2}\mathbf n_{AB}
+O(c^{-2}),
\end{equation}
with
\begin{equation}\label{eq:G}
G \equiv \frac{G_N}{(1+\sigma_1)(1+\sigma_2)}.
\end{equation}
With the definition of active mass in Eq.~(\ref{eq:active_mass}), the Newtonian acceleration matches the GR result, albeit with a rescaled gravitational constant $\mathcal{G}$. Thus, the Newtonian interaction preserves the inverse-square law but with a body-dependent effective gravitational constant, signaling a violation of the SEP. 

Going beyond Newtonian order, additional 1PN terms appear, and the acceleration of body $A$ can be decomposed into three distinct contributions:
\begin{equation}
\mathbf{a}_A \equiv \frac{d\mathbf v_A}{dt}
=
\mathbf{a}_A^{\text{Newton}} + \mathbf{a}_{A}^{L} + \mathbf{a}_{A}^{PF}.
\end{equation}
where the first term is the Newtonian acceleration appearing in Eq.~(\ref{eq:newtonian}). The second term $\mathbf{a}_{A}^{L}$ is the local part of the 1PN acceleration and contains terms that do not depend on the system's velocity relative to the \ae ther. The term $\mathbf{a}_{A}^{L}$ has a GR counterpart, but its expression is modified by the theory's coupling constants. Finally, the third term $\mathbf{a}_A^{PF}$ accounts for the violation of Local Lorentz Invariance. It depends on the velocity $\bm{w}$ of the binary's center-of-mass relative to the \ae ther, including terms proportional to $|\bm w|^2$, $\mathbf v_{21}\cdot\bm{w}$, and $(\mathbf n\!\cdot\!\bm w)^2$. These pieces are governed primarily by the preferred-frame parameters $\alpha_1$ and $\alpha_2$ and introduce secular changes in the orbital elements that are absent in GR. The relative acceleration between the two bodies, retaining only the 1PN contributions, is
\begin{equation}
    \delta \mathbf{a} \equiv \left. \frac{d\mathbf v_A}{dt} - \frac{d\mathbf v_B}{dt}\right|_{1PN} = \mathbf{a}_L + \mathbf{a}_{PF}.
\end{equation}
The first term reads
\begin{align}
\bm{a}_{\rm L}=\frac{m}{r^2}\left[\bm{n}\left(\hat{A}_1 v_{21}^2 +\hat{A}_2 \dot{r}^2+\hat{A}_3 \frac{m}{r}\right)+\dot{r}\hat{B}\bm{v}_{21}\right],
\end{align}
where we have defined
\begin{align}
&\begin{aligned} \hat{A}_1 = \frac{1}{2} \big[ &\mathcal{G}(1-6\eta)-3\mathcal{B}_+-3\Delta\mathcal{B}_- \\
&-\eta(\mathcal{C}_{12}+2\mathcal{E})+\mathcal{G}\mathcal{A}^{(3)} \big], \end{aligned} \label{label_A1}\\
&\hat{A}_2=\frac{3\eta}{2}(\mathcal{G}+\mathcal{E}),\\
&\hat{A}_3=\mathcal{D}+\mathcal{G}\left[2\eta \mathcal{G}+3 \mathcal{B}_++\eta (\mathcal{C}_{12}+\mathcal{E})+3\Delta \mathcal{B}_-\right],\\
&\hat{B}=\mathcal{G}(1-2\eta)+3\mathcal{B}_++3\Delta \mathcal{B}_-+\eta\mathcal{G}+\mathcal{G}\mathcal{A}^{(3)},
\end{align}
while the second term is 
\begin{equation}
\begin{aligned}
&\mathbf{a}_{\text{PF}} = \frac{m}{r^2} \left\{ -\mathbf{n} \left[ \left( \frac{\hat{\alpha}_1}{2} + 2\mathcal{G}\mathcal{A}^{(2)} \right) (\boldsymbol{w} \cdot \mathbf{v}_{21}) \right] \right. \\
&\quad - \mathbf{n} \left[ \frac{3}{2} \left( \hat{\alpha}_2 + \mathcal{G}\mathcal{A}^{(1)} \right) (\boldsymbol{w} \cdot \mathbf{n})^2 \right]  \\
&\quad - \boldsymbol{w} \left[ \frac{\hat{\alpha}_1}{2} (\mathbf{n} \cdot \mathbf{v}_{21}) + \hat{\alpha}_2 (\mathbf{n} \cdot \boldsymbol{w}) \right] \\
&\quad \left. + \mathcal{G}\mathcal{A}^{(2)} \mathbf{v}_{21} (\mathbf{n} \cdot \boldsymbol{w}) - \frac{m |\bm{w}|^2}{2r^2} \left( \mathcal{C}_{12} + \mathcal{G}\mathcal{A}^{(1)} \right) \mathbf{n} \right\}
\end{aligned}
\label{eq:apf}
\end{equation}
where we have defined the following constants:
\begin{align}
m=m_1+m_2, \quad \eta=\frac{m_1m_2}{m^2},\quad \Delta=\frac{m_2-m_1}{m},
\end{align}
and the functions of the sensitivities
\begin{equation}
\begin{aligned}
&\mathcal{G}=\mathcal{G}_{12}, \quad \mathcal{B}_+=\mathcal{B}_{(12)}, \quad \mathcal{B}_-=\mathcal{B}_{[12]},\\
& \mathcal{E}=\mathcal{E}_{12},\quad \mathcal{D}=\frac{m_2}{m}\mathcal{D}_{122}+\frac{m_1}{m}\mathcal{D}_{211},\\  
& \mathcal{A}^{(n)}=\left(\frac{m_2}{m}\right)^n \mathcal{A}_1-\left(-\frac{m_1}{m}\right)^n \mathcal{A}_2.
\end{aligned}
\label{eq:calligraphic}
\end{equation}
The calligraphic objects $\mathcal{G}_{AB}$, $\mathcal{B}_{(AB)}$, $\mathcal{D}_{ABB}$, $\mathcal{E}_{AB}$ and $\mathcal{A}_A$ depend on sensitivities, $\alpha_{1,2}$ and the theory coupling constants (for the exact expression of these quantities, see Eq.~(21) in~\cite{Gupta_2021}). The dependence on the first derivative of the sensitivity enters through $\mathcal{A}_i=-\sigma'_i/(1+\sigma_i)$.
In Eq.~\eqref{eq:apf} we have also introduced the strong-field parameters $\hat{\alpha}_1$ and $\hat{\alpha}_2$ which characterize the preferred-frame effects for compact bodies and are given by

\begin{align}
\hat{\alpha}_1 &= \Delta(\mathcal{C}_{12} + \mathcal{E}) - 6\mathcal{B}_- - 2\mathcal{G}\mathcal{A}^{(2)} \label{eq:alpha_strong1}\\
\hat{\alpha}_2 &= \mathcal{E} - \mathcal{G}\mathcal{A}^{(1)}. \label{eq:alpha_strong2}
\end{align}
They are the strong-field counterpart of the parameters $\alpha_1$ and $\alpha_2$ and become directly proportional to them only when the sensitivities vanish.

The secular evolution of the osculating Keplerian elements — the semi-major axis $a$, eccentricity $e$, inclination $i$, longitude of the ascending node $\Omega$, and argument of periastron $\omega$, which parametrize the size, shape, and orientation of the orbit — under the perturbing acceleration $\delta\mathbf{a}$ follows from the standard Gauss planetary equations (see e.g. Ref.~\cite{Will_2018}). In the absence of perturbations, these elements are constants of motion; the perturbation $\delta\mathbf{a}$  drives their slow secular drift. We decompose $\delta\mathbf{a}$ along the radial, in-plane-tangential and out-of-plane directions of the orbit and average over one period using the modified Keplerian relation with the gravitational constant $\mathcal{G}$ defined in Eq.~\eqref{eq:G}. The full element-by-element analysis is carried out in detail in~\cite{Gupta_2021}; here we summarize the qualitative outcome. The local 1PN sector closely mirrors GR: $\langle\dot a\rangle_L = \langle\dot e\rangle_L = \langle\dot i\rangle_L = \langle\dot\Omega\rangle_L = 0$, while $\langle\dot\omega\rangle_L \neq 0$ produces the standard relativistic periastron advance with Einstein-\ae ther-modified coefficients. The preferred-frame sector is qualitatively new: for a binary moving with velocity $\mathbf w$ relative to the \ae ther, $\langle\dot a\rangle_{PF}$, $\langle\dot e\rangle_{PF}$, $\langle\dot i\rangle_{PF}$ and $\langle\dot\Omega\rangle_{PF}$ are all generically nonzero, and the periastron advance $\langle\dot\omega\rangle = \langle\dot\omega\rangle_L + \langle\dot\omega\rangle_{PF}$ acquires an additional orientation-dependent preferred-frame contribution $\langle\dot\omega\rangle_{PF}$. Closed-form expressions for the post-Keplerian observables required by our timing analysis are collected in Sec.~\ref{sec:post-Keplerian}.

\subsection{Dissipative 1PN Dynamics}
Dissipative dynamics regulate how a binary system loses energy and angular momentum, which causes the orbit to shrink and the orbital period to decay.  A hallmark of Einstein-\ae ther theory is the excitation of additional gravitational wave modes (besides the tensor graviton found in GR). These extra modes carry energy away from the system. In particular, the theory predicts the emission of dipole radiation which  is forbidden in GR. Even the standard quadrupolar emission is modified  due to the extra modes and changes in the propagation speed of gravitons~\cite{Yagi_2014}. An important aspect of the theory is that both conservative and dissipative dynamics depend critically on the sensitivities of the stars. 
The energy flux can be directly related to the change in the binding energy
\begin{equation}
\dot{E}_b = -\mathcal{F}.
\label{eq:balance}
\end{equation}
At 1PN order, the correction to the binding energy and the binary period is such that one still has
\begin{equation}
\frac{\dot{P}_b}{P_b} = -\frac{3}{2}\frac{\dot{E}_b}{E_b}.
\end{equation}

Following~\cite{Yagi_2014}, this rate is expressed as:
\begin{equation}
\label{eq:PbDot_ae}
\begin{split}
\frac{\dot{P}_b}{P_b} = -\frac{3aG}{\mathcal{G}\mu m} \bigg\langle
&\sum_{n=1}^3 \left( \frac{\mathcal{A}_n}{5} \dddot{Q}_{ij} \dddot{Q}_{ij} + 
\mathcal{B}_n \dddot{\mathcal{I}} \dddot{\mathcal{I}} \right) \\
&+ \mathcal{C} \dot{\Sigma}_i \dot{\Sigma}_i + \mathcal{D} \dot{V}_{ij} 
\dot{V}_{ij} \bigg\rangle,
\end{split}
\end{equation}
where $a$ is the semi-major axis, $m$ is the total active gravitational mass, $\mu$ is the reduced mass, and $G$ is the bare gravitational constant. The dimensionless flux coefficients $\mathcal{A}_n$ and $\mathcal{B}_n$ regulate the quadrupole and monopole (breathing) radiation for the tensor ($n=1$), vector ($n=2$), and scalar ($n=3$) modes. The coefficient $\mathcal{C}$ dictates the dipole radiation strength, which is proportional to the dipole moment $\Sigma_i$ and depends on the difference in sensitivities $(s_1 - s_2)$. The coefficient 
$\mathcal{D}$ represents the vector-quadrupole radiation associated with the antisymmetric moment $V_{ij}$. In this expression, $Q_{ij}$ is the mass 
quadrupole moment, $\mathcal{I}$ is its trace, and overdots signify time  derivatives, with the brackets $\langle \dots \rangle$ representing an  average over one orbital period. The final, full expression for this quantity will be given in the next section.

Recently, in Refs.~\cite{Taherasghari:2023rwn, Taherasghari:2025mlf}, the orbital period decay 
was computed by explicitly evaluating the radiation-reaction terms entering the 2.5PN equations 
of motion, obtaining results that differ slightly from those derived via the 
energy-balance relation~\eqref{eq:balance} and worked out in~\cite{Foster:2006az, Foster:2007gr, Yagi_2014}. As discussed in detail in 
Sec.~V.A. of~\cite{Taherasghari:2025mlf}, the discrepancy affects both the dipole and 
quadrupole contributions, and manifests as a difference in the overall prefactor 
multiplying the dipole term in $\dot{E}_b$, while the rest of the expression 
remains unchanged. Specifically, the energy-balance approach yields
\begin{equation}
    \dot{E}_b \propto \left(\frac{2}{c_a v_L^3} + \frac{4(2 - c_a)}{c_\omega v_T}\right),
\end{equation}
whereas the direct PN calculation gives
\begin{equation}\label{directPN}
    \dot{E}_b \propto \left(1 + \frac{6(2 - c_a)}{c_\omega v_T}\right),
\end{equation}
with the propagation speeds given by
\begin{equation}
    v_T^2 = \frac{c_\omega}{2c_a}, \qquad
    v_L^2 = \frac{(2 - c_a)c_\theta}{3(2 + c_\theta)c_a}.
\end{equation}
The most notable structural difference is that the direct PN result lacks any dependence 
on $v_L$, the propagation speed of the spin-0 (longitudinal scalar) mode. This is 
somewhat surprising: since the spin-0 mode carries energy away from the system and 
is expected to remain physical at all PN orders~\cite{Jacobson:2004ts,Garfinkle:2007bk,Franchini:2021bpt}, one would naturally expect its 
propagation speed to appear in the flux. We also note that the direct PN prefactor~\eqref{directPN} does not 
vanish in the limit $v_T, v_L \to \infty$, whereas the energy-balance expression does.
The latter appears to be the physically expected behavior, as it reduces to the GR limit (i.e. no dipole flux) when
the fields become non-dynamical.\footnote{One may argue that the dipole flux
goes to zero in the GR limit in the direct-PN case as well, because the sensitivities  approach zero for a pulsar in the GR limit. However, the GR limit should be recovered \textit{also} for a hypothetical matter source with constant sensitivities (because the sensitivities couple matter to the \ae ther, which becomes non-dynamical).
}
For these reasons, in this work we 
follow the expression adopted in the literature prior to~\cite{Taherasghari:2025mlf}, and defer a thorough investigation of this discrepancy to a future 
study~\cite{Aether_collapse}. We nonetheless point out that our framework requires only minimal 
modifications to instead accommodate the direct-PN expression.

\subsection{Explicit expressions for the post-Keplerian parameters}
\label{sec:post-Keplerian}
In this section, we find the post-Keplerian parameters in Einstein-\ae ther which we will use below to obtain constraints on the theory variables by using the posteriors of the timing  analysis. These parameters are: the Einstein delay parameter $\gamma$, the Shapiro delay parameters $r$ and $s$, the periastron advance rate $\dot{\omega}$ and the orbital period derivative $\dot{P}_b$. 

The Einstein delay results from the effects of both special relativity (time dilation due to relative motion) and general relativity (gravitational redshift due to the presence of a gravitational field). Using the usual post-Newtonian coordinates and the 2-body metric at 1PN order, and following the same computation as in standard GR, one finds 
\begin{equation}
\begin{split}
&\gamma =  e \left( \frac{P_b}{2 \pi} \right)^{1/3} \left( \frac{G_N^{2/3}}{c^2} \right) (1+\sigma_p)^{1/3} (1+\sigma_c)^{1/3}  \\
&\times \left[ \frac{2\tilde{m}_c^2 +\tilde{m}_c\tilde{m}_p}{(\tilde{m}_c+\tilde{m}_p)(m_c+m_p)^{1/3}} \right]
 \; .
\end{split}
\end{equation}
where $m_p$ and $m_c$ are the masses of the pulsar and the companion, respectively, and we have used the Keplerian relation $P_b=2\pi\sqrt{a^3/(\mathcal{G}M)}$. The Einstein delay will simply be $\Delta_E= \gamma \sin{u}$ where $u$ is the eccentric anomaly. Therefore, the delay is formally equivalent to the classical post-Keplerian expression in GR, but $\gamma$ gains contributions from sensitivity terms.

The periastron advance rate, $\dot{\omega}$, is, together with the orbital period derivative, one of the best-measured post-Keplerian parameters in binary pulsars. In Einstein-\ae ther, the accumulated angle per orbit was derived in~\cite{Will_2018}, where it appears as $\Delta \omega = \Delta \tilde{\omega} - \cos{i}\Delta \Omega$, with  $\tilde{\omega}$  a redefinition of $\omega$, $i$ the inclination of the orbital plane and $\Omega$ the longitude
of the ascending node. Defining $p\equiv a(1-e^2)$ as the semi-latus rectum and combining Eqs.~(55) of~\cite{Will_2018}, we obtain  

\begin{equation}\label{eq:deltaw}
 \Delta \omega = \Delta \omega^{(L)} + \Delta \omega^{(PF)}   
\end{equation}
where $\Delta \omega^{(L)}$ is given by Eq.~(A.18) in~\cite{Gupta_2021}. We report explicitly the preferred-frame contribution $\Delta \omega^{(PF)}$:
\begin{equation}\label{eq:16}
\begin{split}
  &\Delta \omega^{(PF)} =  \hat{\alpha}_1 \pi  \left(\frac{\tilde{m}}{\mathcal{G}p} \right)^{1/2} F(e) \\
  & \times\left[\frac{w_h}{\tan{i}} e \cos{\omega}-\frac{w_q}{e}\sqrt{1-e^2}+\frac{w_q}{e F(e)} \right] + \hat{\alpha}_2 \pi F(e) \\
  & \times \left[ \frac{2 w_h (w_p \sin \omega + w_q \sqrt{1 - e^2} \cos \omega)}{\mathcal{G} \tan{i} \sqrt{1-e^2}}   - w_p^2 w_q^2 F(e) \right]\\
  &+ \frac{4 \pi w_q \mathcal{A}^{(2)}\mathcal{G}}{e} \left(\frac{\tilde{m}}{\mathcal{G}p}\right)^{1/2}
 \end{split}
\end{equation}
with $m=m_p+m_c$ the total mass, $F(e) = 1/(1+\sqrt{1-e^2})$
and $w_p, w_q$ and $w_h$ the components of $\bm{w}$ in the orbital frame $\{\bm{e}_p,\bm{e}_q,\bm{e}_h\}$, where $\bm{e}_p$ is the direction of the pericenter, $\bm{e}_h$ is aligned with the angular momentum and $\bm{e}_q=\bm{e}_h\times\bm{e}_p$.

Notice that in the second line of Eq.~(\ref{eq:16}) $\omega$ is not known; indeed, the pericenter trend can be written as $\omega=\omega_0 + \dot{\omega}t$ (where $t$ is the observation time) but $\dot{\omega}=\Delta \omega /P_b$. However, in first approximation, we can assume  $ \omega \simeq \omega_0 + (\Delta \omega^{(L)}/P_b)t$ and since we just need a zero-order quantity, we will consider $\omega \approx \omega_0$  in the second line of  Eq.~(\ref{eq:16}). Since pulsar timing gives access to the rate $\dot{\omega} = \Delta\omega / P_b$, dividing Eq.~(\ref{eq:deltaw}) by $P_b$ and substituting, the periastron advance reads
\begin{widetext}
\begin{equation}
\begin{split}
 \dot{\omega} =  \frac{6\pi m}{{\cal G}p P_b}&\left[{\cal G}{\cal B}_++\frac{1}{6}\left({\cal G}^2-{\cal D}\right)+\frac{1}{6}{\cal G}\left(6 \Delta {\cal B}_- + \eta (2{\cal C}_{12}+{\cal E})+{\cal G}{\cal A}^{(3)}\right)\right]\\
 & + \hat{\alpha}_1 \pi  \left(\frac{\tilde{m}}{\mathcal{G}p} \right)^{1/2} \frac{F(e)}{P_b} \left[\frac{w_h}{\tan{i}} e \cos{\omega_0}-\frac{w_q}{e}\sqrt{1-e^2} +\frac{w_q}{e F(e)} \right] \\
 &  +  \hat{\alpha}_2 \pi \frac{F(e)}{P_b} \left[ \frac{2 w_h (w_p \sin \omega_0 + w_q \sqrt{1 - e^2} \cos \omega_0)}{\mathcal{G} \tan{i} \sqrt{1-e^2}}   - w_p^2 w_q^2 F(e) \right]+ \frac{4 \pi w_q \mathcal{A}^{(2)}\mathcal{G}}{e P_b} \left(\frac{\tilde{m}}{\mathcal{G}p}\right)^{1/2}
\end{split}
\end{equation}
\end{widetext}
where we remember that $\hat{\alpha}_{1,2}$ depend on sensitivities (and their derivatives) and are given by Eqs.~\eqref{eq:alpha_strong1}-\eqref{eq:alpha_strong2}.

The Shapiro delay is a relativistic effect whereby the propagation time of 
electromagnetic signals is increased by the spacetime curvature produced by a 
massive body. In pulsar timing, it manifests as a periodic delay in pulse arrival 
times as the signal passes near the companion. The Shapiro delay in Einstein--\ae ther 
theory has been studied in the Solar System context in~\cite{Shapiro_Solar_2020}, 
building on~\cite{BH_in_EA_charged}; however, for compact objects one must also 
account for the bodies' sensitivities. A direct computation using the binary 1PN 
metric in Eq.~\eqref{eq:g0i_1PN} yields a result of the same functional form as in GR,
\begin{equation}
    \Delta_S = -2r \ln \left\{ 1 - e \cos u - s f(u) \right\}
\end{equation}
where 
\begin{equation}
f(u)=\left[ \sin \omega (\cos u - e) + \sqrt{1 - e^2} \cos \omega \sin u \right]
\end{equation}
and $r$ (range) and $s$ (shape) are given by $r\equiv G_N \tilde{m}_c$ and $s\equiv\sin{\iota}$. These expressions are therefore identical to the GR ones. However, the shape 
parameter $s = \sin i$ is related to the component masses through the binary mass 
function, which is based on Kepler's third law. Since the latter is modified in 
Einstein--\ae ther theory~\cite{Yagi_2014}, this relation acquires a beyond-GR 
correction. Indeed, using the modified Kepler's third law, the binary mass function 
at Newtonian order reads
\begin{equation}\label{eq:s}
 \left( \frac{2 \pi}{P_b}   \right)^{2/3}  \left( \frac{x \tilde{m} c}{\tilde{m}_c} \right) (\mathcal{G} m)^{-\frac{1}{3}} = \sin{i} 
\end{equation}
where $x\doteq a \sin{i} /c$ is the projected semi-major axis of the pulsar. Notice that $s,P_b$ and $x$ appearing in the above equation are all parameters that enter the pulsar timing model and that the only difference with the GR/weak-field case is the presence of the active total mass $m$ and the modified gravitational constant $\mathcal{G}$. 

Finally, we turn to the orbital period derivative. Evaluating the average in 
Eq.~\eqref{eq:PbDot_ae} (the details of which are given in 
Appendix~\ref{app:average}) we obtain for an eccentric binary with masses $m_1$, $m_2$
\begin{widetext}
\begin{equation}
\begin{split}
\frac{\dot P_b}{P_b} &=
  -\frac{3\,m_1 m_2\,(1+\alpha_1/8)}{(2\pi)^{2/3}\,m^{1/3}}
  \Biggl\{
\\[4pt]
&\quad
  \frac{2^{1/3}\pi^{10/3}\bigl[(1-s_1)(1-s_2)\bigr]^{2/3}}
       {(1-e^2)^{7/2}\,m^{2/3}\,P_b^{8/3}}
  \Biggl[
    \frac{64}{15}\!\left(12
      - \frac{\mathcal{K}}{c_\omega^3 m^2\alpha_1^4(8+\alpha_1)}
        \sqrt{\frac{\alpha_1}{\alpha_1-8\alpha_2}}
      \right)
\\&\qquad
    + e^2\!\left(
        \frac{2336}{15}
        - \frac{64\,\Delta_m^2}{9\,m^2\alpha_1^3(8+\alpha_1)}
          \sqrt{\frac{\alpha_1-8\alpha_2}{\alpha_1}}
        - \frac{584\,\mathcal{K}}{45\,c_\omega^3 m^2\alpha_1^4(8+\alpha_1)}
          \sqrt{\frac{\alpha_1}{\alpha_1-8\alpha_2}}
      \right)
\\&\qquad
    + \frac{2e^4}{45}\!\left(444
        - \frac{40\,\Delta_m^2}{m^2\alpha_1^3(8+\alpha_1)}
          \sqrt{\frac{\alpha_1-8\alpha_2}{\alpha_1}}
        - \frac{37\,\mathcal{K}}{c_\omega^3 m^2\alpha_1^4(8+\alpha_1)}
          \sqrt{\frac{\alpha_1}{\alpha_1-8\alpha_2}}
      \right)
  \Biggr]
\\[4pt]
&\quad
  -\,\frac{2^{2/3}\pi^{8/3}(s_1-s_2)^2
           \bigl[(4+3e^2)w_p^2+(4+e^2)w_q^2\bigr]}
          {15\,(1-e^2)^{5/2}\,m^{4/3}\,P_b^{2}\,\alpha_1^2}
  \left(
    \frac{\sqrt{2}\,c_\omega}{\bigl(-c_\omega/\alpha_1\bigr)^{7/2}}
    - \frac{576\,\alpha_1}{8+\alpha_1}
      \left(\frac{\alpha_1-8\alpha_2}{\alpha_1}\right)^{\!5/2}
  \right)
\\[4pt]
&\quad
  +\,\frac{32\cdot2^{2/3}\pi^{8/3}(s_1-s_2)^2(2+e^2)\,|\mathbf{w}|^2}
          {15\,(1-e^2)^{5/2}\,m^{4/3}\,P_b^{2}\,\alpha_1^4}
  \left(
    \frac{\sqrt{2}\,\sqrt{-c_\omega/\alpha_1}\;\alpha_1^6}{c_\omega^3}
    - \frac{36\,(\alpha_1-8\alpha_2)^3}{8+\alpha_1}
      \sqrt{\frac{\alpha_1}{\alpha_1-8\alpha_2}}
  \right)
\\[4pt]
&\quad
  -\,\frac{8e(4+e^2)\pi^3(s_1-s_2)\bigl[(1-s_1)(1-s_2)\bigr]^{1/3}w_q}
          {(1-e^2)^3\,m^2\,P_b^{7/3}\,\alpha_1^2}
  \left(
    \frac{\sqrt{2}\,\bar{\mathcal{S}}\,\sqrt{-c_\omega/\alpha_1}\;\alpha_1^4}
         {c_\omega^3}
    - \frac{8\,\Delta_m^{(6)}}{8+\alpha_1}
      \left(\frac{\alpha_1-8\alpha_2}{\alpha_1}\right)^{\!3/2}
  \right)
\\[4pt]
&\quad
  +\,\frac{4\,(s_1-s_2)^2}{3\,\alpha_1}
    \left(
      \frac{\sqrt{2}}{\bigl(-c_\omega/\alpha_1\bigr)^{3/2}}
      - \frac{16}{8+\alpha_1}
        \left(\frac{\alpha_1-8\alpha_2}{\alpha_1}\right)^{\!3/2}
    \right)
\\&\qquad\qquad\times
    \left[
      \frac{2\cdot2^{2/3}(2+e^2)\pi^{8/3}}
           {(1-e^2)^{5/2}\,m^{4/3}\,P_b^{2}}
      + \frac{\pi^{10/3}\!\left(\mathcal{P}(e)\,\delta a_{\rm sec}
              + \mathcal{Q}(e)\,\delta e\right)}
             {8\cdot2^{2/3}\,m^{5/3}\,P_b^{7/3}}
    \right]
  \Biggr\},
\end{split}
\label{eq:pbdot_avg}
\end{equation}
\end{widetext}
where $m \equiv m_1+m_2$,
\begin{equation}
  \bar{\mathcal{S}} \equiv m_2 s_1 + m_1 s_2,
\end{equation}
\begin{align}
  \Delta_m &\equiv
    \alpha_1\bigl[m_2(2s_1+\alpha_1)+m_1(2s_2+\alpha_1)\bigr]
    \notag\\&\quad
    - 2\alpha_2\bigl[m_2(8s_1+\alpha_1)+m_1(8s_2+\alpha_1)\bigr],
\end{align}
with $\Delta_m^{(6)}$denoting the same combination with the coefficient $2$ 
replaced by $6$ in front of $s_{1,2}$,
\begin{align}
 \mathcal{K} &\equiv
    2c_\omega^3\bigl[m_2(8s_1+\alpha_1)+m_1(8s_2+\alpha_1)\bigr]^2
    (\alpha_1-8\alpha_2)^3
    \notag\\&
    + 3\sqrt{2}\,\bar{\mathcal{S}}^2\,\alpha_1^5(8+\alpha_1)
      \sqrt{-c_\omega(\alpha_1-8\alpha_2)},
\end{align}
and the eccentricity polynomials
\begin{align}
  &\mathcal{P}(e) = -256
    + e^2\Bigl[
        2e^2\bigl(256 + 960e^2 + 70e^4(32+60e^2 \notag\\
    &+99e^4)\bigr) - 10\bigl(128 + 21e^2(16+32e^2+55e^4)\bigr)
      \Bigr],
\end{align}
\begin{align}
  \mathcal{Q}(e) &=
    \frac{1}{2}\!\left(\frac{m P_b}{2\pi}\right)^{\!1/3}\!\!\Bigl[
      512e
      \notag\\&\quad
      + e^3\bigl(105(32+96e^2+220e^4+429e^6)
      \notag\\&\quad
      - 5(288+7e^2(160+420e^2+891e^4))\bigr)
    \Bigr].
\end{align}

For PSR J1738+0333, $e\sim \mathcal{O}(10^{-7})$ (Table~\ref{tab:timing-results}), so the eccentricity-dependent terms in Eq.~\eqref{eq:pbdot_avg} contribute at fractional level $\lesssim e^2 \sim 10^{-13}$ relative to the leading term, far below the measurement precision; they are retained only for completeness and applicability to more eccentric systems.

An important remark concerns the measured value of $\dot{P}_b$, which is 
contaminated by kinematic effects, namely the Shklovskii effect and the 
differential Galactic acceleration~\cite{5307C}. The corrected orbital period derivative 
is given by
\begin{equation}
    \dot{P}_b^\mathrm{corr} = \dot{P}_b - \dot{P}_\mathrm{Shk} - \dot{P}_\mathrm{Gal},
    \label{eq:pbdot_corr}
\end{equation}
where
\begin{equation}
    \dot{P}_\mathrm{Shk} = \frac{P_b\, \mu^2 d}{c}\,, 
    \qquad 
    \dot{P}_\mathrm{Gal} = \frac{P_b\, a_\mathrm{Gal}}{c}\,,
    \label{eq:pbdot_kin}
\end{equation}
with $\mu^2 = \mu_\alpha^2 + \mu_\delta^2$ the total proper motion squared ($\mu_\alpha$ and $\mu_\delta$ correspond to the timing parameters 
\texttt{PMRA} and \texttt{PMDEC} respectively), $d$ the distance to the 
pulsar, and $a_\mathrm{Gal}$ the differential Galactic acceleration at 
the position of the binary~\cite{accelerations_effects}.\footnote{In 
principle, additional terms $\dot{P}_b^{\dot{m}}$ and $\dot{P}_b^{T}$, 
arising from mass loss and tidal deformation of the companion star 
respectively, could appear in Eq.~\eqref{eq:pbdot_corr}. However, they are generally negligible for this system and 
we do not consider them here.} We note that in some references 
$\dot{P}_b^\mathrm{corr}$ is denoted $\dot{P}_b^\mathrm{GW}$, as it 
represents the orbital decay due to gravitational-wave emission alone.

Both kinematic corrections are proportional to $P_b$ and depend on the 
distance through the respective accelerations, with $a_\mathrm{Gal}$ 
further depending on the pulsar sky position and the Galactic potential 
model (see Appendix~\ref{app:pbdot_gal} for details). We can therefore write compactly for the observed period derivative
\begin{equation}
    \dot{P}_b= \dot{P}_b^\mathrm{corr}  + 
    f\!\left(P_b,\,\mu_\alpha,\,\mu_\delta,\,\alpha,\,\delta,\,\Xi\right),
    \label{eq:pbdot_f}
\end{equation}
where $\Xi$ denotes the parameters of the Galactic potential model and $\dot{P}_b^\mathrm{corr}$ is obtained from Eq.~\eqref{eq:pbdot_avg}.

\section{The pulsar - white dwarf system PSR J1738+0333}
\label{sec:obs_and_dataset}

PSR\,J1738+0333 is a millisecond pulsar ($P=5.9$\,ms) in a short period ($P_b=8.5$\,hours) and nearly circular ($e=3.4\times10^{-7}$) orbit around a low-mass WD companion~\cite{Freire:2007sb, Antoniadis:2012vy}. The observed time derivative of the orbital period $\dot{P}_b$ is large ($\mathcal{O}(10^{-14})$\,s\,s$^{-1}$) and has been determined with great precision. The high precision measurements of the system distance, through VLBI observations~\cite{VLBI_2023}, and of the proper motion~\cite{Freire:2012} allow for the determination of the contributions to the measured $\dot{P}_b$ that are due to the real acceleration imparted to this system by the gravitational potential of the Milky Way, and to the apparent acceleration due to the transverse motion, a.k.a. Shklovskii effect, as seen from the observer (see Eq.\eqref{eq:pbdot_corr}). Optical observations of the WD companion yielded a measurement of its 
mass~\cite{Antoniadis:2012vy}, $M_c = 0.18\,M_\odot$, which combined with the 
mass function gives an estimate of the pulsar mass, $m_p = 1.42\,M_\odot$ (see Appendix~\ref{app:corner}).

PSR\,J1738+0333 is also characterized by a remarkable timing stability and precision ($\sim 2\mu$s), which led to its inclusion into pulsar timing array programs such as the European Pulsar Timing Array (EPTA)~\cite{EPTA:2016ndq} and the North American Nanohertz Observatory for Gravitational waves (NANOGrav)~\cite{NANOGrav:2023gor} collaborations. Thanks to all of these properties, PSR\,J1738+0333 has provided some of the best constraints on scalar-tensor theories ~\cite{BenSalem:2023xgy} and quadratic scalar-tensor gravity, has set the best limit on dipole radiation, and has also excluded TeVeS-like theories~\cite{Freire:2012}.

\subsection{Observations and datasets}

The pulsar timing technique consists of modeling the times of arrival (ToAs) of pulses
at the telescope as a deterministic function of a set of timing parameters
$\vec{\lambda}$, which includes astrometric, spin, dispersion, and orbital
parameters. The timing model predicts the phase of each pulse, and the
residuals $\vec{r}(\vec{\lambda}) = \vec{t}_\mathrm{obs} - \vec{t}_\mathrm{model}(\vec{\lambda})$
encode the mismatch between the data and the model. ToAs are extracted by cross-correlating each 
observed pulse profile against a high-quality noise-free template, known as 
the standard profile~\cite{Taylor:1992kea}. For each dataset, the standard profile is 
constructed by summing all observations above a minimum signal-to-noise 
threshold, and then averaging over frequency 
channels and time. The time and frequency resolution 
adopted for each telescope reflects a trade-off between retaining 
sensitivity to profile evolution and maximising ToA precision.
\\
\begin{figure*}[htbp]
    \centering
    \includegraphics[width=\textwidth]{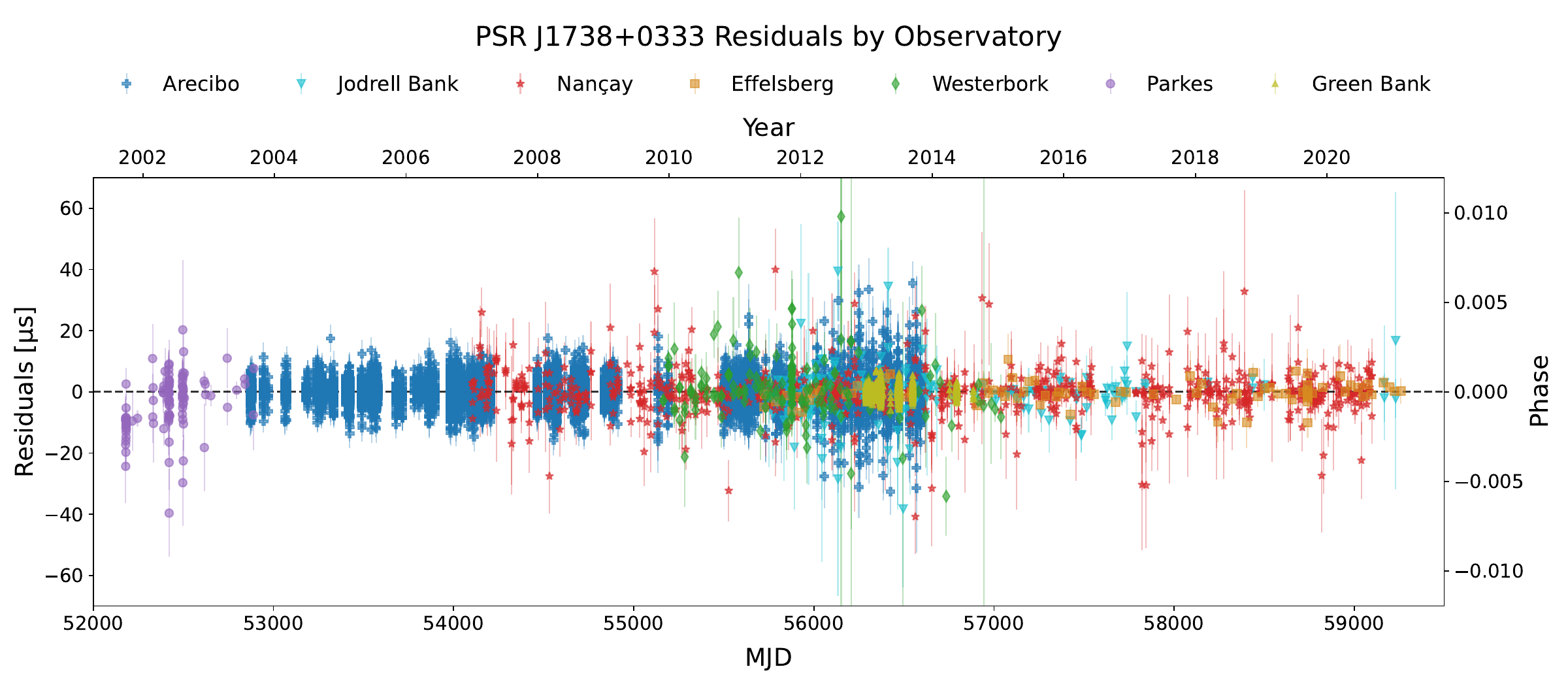}
    \caption{Timing residuals with respect to the posterior median timing model, with the noise realization subtracted, for PSR J1738+0333 grouped by observatory. A consistent picture is obtained using the maximum-posterior estimator. The plot shows multi-decade narrowband ToAs from eight different telescopes.}
    \label{fig:residuals_by_obs}
\end{figure*}
In this work, we analyzed the ToAs  of PSR\,J1738+0333 from the EPTA second data release~\cite{EPTA-DR2} and from the NANOGrav 9-year data release~\cite{NANOGRAV-9Y}. In addition, we also used the ToAs of the previous similar analysis by Freire et al.~\cite{Freire:2012} and further observations taken with the Green Bank Telescope. In order to further increase the precision of the measurement of the orbital parameters, two intensive campaigns with the Arecibo and Effelsberg telescopes were conducted over nearly one year, between July 2019 and June 2020~\cite{BenSalem:2023xgy}.

Overall, we collected a total of 25,054 narrowband ToAs along an epoch range of $\sim20$\,years from 27 September 2001 to 14 February 2021. The majority of ToAs come from L-band observations, while a minority come from S-band Arecibo and P-band Westerbork Synthesis Radio Telescope (WSRT) observations. A summary of the ToA sets is presented in Table~\ref{tab:toa-summary}.
\begin{table}[ht]
    \centering
    \begin{tabular}{lcccc}
    \hline
        \hline
        Telescope & Frequency band & MJD range & \#ToAs \\
        \hline
        \multicolumn{4}{c}{EPTA DR2} \\
        \hline
Effelsberg & L-band & 55723-59383 & 87 \\
Jodrell Bank & L-band & 55734-59230 &  209 \\
Nançay & L-band & 54103-59097 & 705 \\
Westerbork & L-band & 55164-57039 & 81 \\
        \hline
\multicolumn{4}{c}{NanoGrav 9yr}\\
        \hline
Arecibo & L-band & 55135-56591 & 1881 \\
Arecibo & S-band & 55135-56591 & 742 \\
        \hline
\multicolumn{4}{c}{ToAs used in Ref.~\cite{Freire:2012}}\\
        \hline
Arecibo &  L-band & 52872-55813 & 17376 \\
Nançay & L-band & 54105-55407 & 50 \\
Parkes & L-band & 52179-52891 & 101 \\ 
Westerbork & L-band & 55164-55962 & 28 \\
Westerbork & P-band &  55193-55962 & 50 \\
        \hline
\multicolumn{4}{c}{New ToAs}\\
        \hline
Green Bank & L-band & 56290-56886 & 3747 \\
        \hline
        \hline
\end{tabular}
    \caption{Summary of the ToAs analysed in this work.}
    \label{tab:toa-summary}
\end{table}

\subsection{Bayesian timing analysis}
\label{sec:bayesian_timing}
In the Bayesian framework~\cite{Lentati:2013rla, Susobhanan:2024yef}, the likelihood of
the ToA data $\vec{t}$ given the timing parameters $\vec{\lambda}$ and
noise parameters $\vec{\beta}$ is given by
\begin{equation}
    \mathcal{L}(\vec{t} \mid \vec{\lambda}, \vec{\beta})
    = \frac{1}{\sqrt{\det(2\pi\,\mathsf{C})}}\,
      \exp\!\left(-\frac{1}{2}\,\vec{r}^{\,\top}
      \mathsf{C}^{-1}\vec{r}\right),
\label{eq:likelihood}
\end{equation}
where $\mathsf{C}$ is the total noise covariance matrix. This matrix
receives contributions from both white and red noise processes.
White noise is modeled through three standard parameters: $E$ or \texttt{EFAC},
a multiplicative factor rescaling the radiometric uncertainties $\sigma_i$;
$Q$ or \texttt{EQUAD}, an additional noise added in quadrature; and
$J$ or \texttt{ECORR}, a term that correlates all ToAs within a single
epoch~\cite{Lentati:2013rla, Susobhanan:2024yef}. The noise covariance is
\begin{equation}
    \mathsf{N}_{ij}  =  \left[\left(E_{\mu(i)}^2 \sigma_i^2 + Q_{\mu(i)}^2\right)\delta_{ij}+ J_\mu^2 \delta_{e(i)e(j)}\right]\delta_{\mu(i)\mu(j)}.
    \label{eq:white}
\end{equation}
where $i=\{1,N_{ToAs}\}$ runs over the ToAs, $\mu$ labels the receiver-backend pair and $e(i)$ is the epoch of the ToA, so that $\delta_{e(i)e(j)}$ correlates ToAs corresponding to the same time epoch but different frequencies. 

Depending on the pulsar, time-correlated (red) noise processes may also
contribute to the residuals. These include spin noise, dispersion measure
(DM) noise, and the stochastic gravitational-wave background. A red noise
process with power spectral density decreasing with (conjugate) frequency
is typically modeled as a truncated Fourier series~\cite{Lentati:2013rla},

\begin{equation}
\begin{split}
    \Delta(t) = {} & \left(\frac{\nu}{\nu_\mathrm{ref}}\right)^{\!\alpha}
    \sum_{j=1}^{N_f}
    \left[ a_j \cos\!\left(2\pi j f_1 (t - t_0)\right) \right. \\ 
    & \left. + \, b_j \sin\!\left(2\pi j f_1 (t - t_0)\right) \right], 
\end{split}
\label{eq:red_noise}
\end{equation}
where $\nu$ is the observing frequency, $\nu_\mathrm{ref}$ is a reference
frequency (typically $\nu_\mathrm{ref}=1400 MHz$), $\alpha$ is the chromatic index ($\alpha = 0$ for achromatic
spin noise, $\alpha = 2$ for DM noise), $N_f$ is the number of harmonics,
$f_1$ is the fundamental frequency ($f_1=1/T_{obs}$), $t_0$ is a fiducial epoch, and $a_j$,
$b_j$ are Fourier coefficients.

In the power-law model, the coefficients $a_j$ and $b_j$ are treated as
zero-mean Gaussian random variables with variance equal to the one-sided
power spectral density evaluated at frequency $f_j = j f_1$,
\begin{equation}
    \sigma_j^2 \equiv P(f_j)
    = \frac{A^2}{12\pi^2 f_\mathrm{yr}^3}\,
      f_1\!\left(\frac{f_j}{f_\mathrm{yr}}\right)^{-\Gamma},
\label{eq:powerlaw}
\end{equation}
where $f_\mathrm{yr} = 1\,\mathrm{yr}^{-1}$ and $A$ and $\Gamma$ are the amplitude and the spectral index, respectively, which are themselves sampled. An analogous model describes DM noise, with $\alpha = 2$ in
Eq.~\eqref{eq:red_noise} and its own amplitude and spectral index.

Rewriting Eq.~\eqref{eq:red_noise} as $\Delta(t)=\sum_{j=1}^{2N_f} F_j w_j$ with $\vec{w}=\{a_1,b_1...a_{N_f},b_{N_f}\}$ the total noise covariance matrix, marginalized over the coefficients $a_j$ and $b_j$, can be written as~\cite{PhysRevD.87.104021}
\begin{equation}
\mathsf{C}=\mathsf{N}+\mathsf{F}\mathsf{\Phi} \mathsf{F}^T
\end{equation}
where $\Phi$ is the $N_f\times N_f$ covariance matrix of the Fourier coefficients $\mathsf{\Phi}_{ij}=\langle w_i,w_j\rangle=\sigma_j^2\delta_{ij}$. The red-noise covariance is thus incorporated into $\mathsf{C}$ in Eq.~\eqref{eq:likelihood}, closing the noise model. 

The joint posterior over timing and noise parameters,
\begin{equation}
    p(\vec{\lambda},\vec{\beta}\mid\vec{t})
    \propto \mathcal{L}(\vec{t}\mid\vec{\theta},\vec{\beta})\,
    \pi(\vec{\theta})\,\pi(\vec{\beta}),
\label{eq:posterior}
\end{equation}
can be sampled using standard Markov chain Monte Carlo methods. The posterior on $\vec{\lambda}$, obtained after marginalization over the noise 
parameters $\vec{\beta} = \{A_\mathrm{red}, \Gamma_\mathrm{red}, A_\mathrm{DM}, 
\Gamma_\mathrm{DM}, E_\mu, Q_\mu, J_\mu\}$, constitutes the input to the resampling procedure 
described in Sec.~\ref{sec:resampling}.

The Bayesian timing analysis was performed using \textsc{Vela}~\cite{Susobhanan:2024yef},
which implements the full non-linear pulsar timing and noise model with efficient parallelization,
provides a \texttt{Python} binding (\texttt{pyvela}) and handles data I/O, clock corrections,
and ephemeris computation via \textsc{pint}~\cite{Luo:2020ksx, Susobhanan:2024gzf}, and is designed to work with both narrowband and wideband analysis paradigms. Within \textsc{Vela}, we employed the ensemble sampler \texttt{emcee}~\cite{Foreman-Mackey:2012any} to draw samples from the likelihood. 
  
\begin{table*}[t]
\centering
\begin{tabular}{lc}
\hline
\hline
\multicolumn{2}{c}{Fit parameters} \\
\hline
Ephemeris version & DE440 \\
Units & TDB \\
Clock & TT(BIPM2021) \\
Reference epoch & 54999.9998161703821147 \\
MJD epoch range & 52179$-$59259 \\
Number of ToAs & 25054 \\
\hline
\multicolumn{2}{c}{Astrometric and rotational parameters} \\
\hline
Right Ascension ($\alpha$, J2000) & 17:38:53.9663730800(5) \\
Declination ($\delta$, J2000) & 3:33:10.87199174(2) \\
Proper motion in Right Ascension ($\mu_\alpha $, mas\,yr$^{-1}$) & 7.081(6) \\
Proper motion in Declination ($\mu_\delta $, mas\,yr$^{-1}$) & 5.056(16) \\
Parallax ($\varpi$, mas) & 0.603(13) \\
Spin frequency ($\nu$, Hz) & 170.9373725375267(3) \\
First derivative of the spin frequency ($\dot{\nu}$, 10$^{-16}$\,Hz\,s$^{-1}$) & -7.04750(5) \\
Dispersion Measure ($DM$,pc\,cm$^{-3}$) & 33.7722(5) \\
First derivative of the Dispersion Measure ($DM1$,pc\,cm$^{-3}$\,yr$^{-1}$) & -0.00070(6) \\
Second derivative of the Dispersion Measure ($DM2$,pc\,cm$^{-3}$\,yr$^{-2}$) & -0.00001(3) \\
\hline
\multicolumn{2}{c}{Orbital parameters} \\
\hline
Binary model & ELL1 \\
Orbital period ($P_b$, days) & 0.354790734359(2) \\
Projected semi$-$major axis ($a$, lt-s) & 0.34342911(3) \\
Time of passage at the ascending node ($T_\mathrm{asc}$, MJD) & 55441.763844396(7) \\
First Lagrange-Laplace parameter ($\eta_{LL}\equiv e\sin\omega$, 10$^{-7}$) & -0.7(17) \\
Second Lagrange-Laplace parameter ($\kappa_{LL}\equiv e\cos\omega$, 10$^{-7}$) & -0.9(17) \\
Measured time derivative of the orbital period ($\dot{P}_b$, 10$^{-14}$\,s\,s$^{-1}$) & --1.82(25) \\
\hline
\multicolumn{2}{c}{Derived parameters} \\
\hline
Spin period ($P$, ms)& 5.85014398237(10) \\
First derivative of the spin period ($\dot{P}$, $10^{-20}$\,s\,s$^{-1}$) & 2.41204(2) \\
Orbital eccentricity ($e$, $10^{-7}$) & 2.2(13) \\
\hline
\hline
\end{tabular}
    \caption{Timing and derived parameters for PSR J1738+0333. Fitted parameters and their uncertainties are reported as the median and half the 68\%
credible interval (16th–84th percentiles) of the marginalized posteriors from a Bayesian analysis performed with \textsc{Vela}. Derived quantities are computed from the posterior medians. A consistent set of parameters is obtained using the MAP estimator. Figures in parentheses denote the uncertainty on the last quoted digit.}
    \label{tab:timing-results}
\end{table*}

\subsection{From timing to theory parameters: the resampling strategy}
\label{sec:resampling}

The Bayesian timing analysis described in the previous section gives a posterior 
distribution in the form of Eq.~\eqref{eq:posterior}, where 
$\vec{\lambda}=\{\vec{\theta},\vec{\zeta}\}$ collects all timing parameters, 
with $\vec{\zeta}$ denoting spin and astrometric parameters and 
$\vec{\theta}=\{\vec{\theta}_\mathrm{orb},\vec{\theta}_\mathrm{PK}\}$ the orbital 
and post-Keplerian parameters. After marginalization over the noise parameters 
$\vec{\beta}$ and $\vec{\zeta}$, we obtain $p(\vec{\theta}\mid\vec{t})$. 

Because 
the orbital eccentricity of PSR~J1738+0333 is small, we adopt the ELL1 timing 
model~\cite{Lange:2001rn}, in which the Laplace--Lagrange parameters 
$\eta_{LL} = e\sin\omega$ and $\kappa_{LL} = e\cos\omega$ replace eccentricity and periastron 
argument as directly fitted quantities, so that
\begin{equation*}
\vec{\theta}_\mathrm{orb} = \{\varpi,\,\mu_\alpha,\,\mu_\delta,\,P_b,\,a_1\sin i,\,\eta_{LL},\,\kappa_{LL},\,T_\mathrm{asc}\},
\end{equation*}
comprising the parallax, proper motions, orbital period, projected semi-major axis, 
Laplace--Lagrange parameters, and time of ascending node. The post-Keplerian observables $\vec{\theta}_\mathrm{PK}$ described in 
Sec.~\ref{sec:post-Keplerian} are $\dot{P}_b$, $\dot{\omega}$, $r$, $\sin i$, and 
$\gamma$. Also because of the small eccentricity, only $\dot{P}_b$ is measurable; 
the Shapiro delay parameters $r$ and $\sin i$ are included in the timing model as 
fixed quantities derived in~\cite{Antoniadis:2012vy}, rather than fitted, to improve 
the timing solution.

To constrain Einstein--\ae ther theory, we ultimately seek the posterior 
$p(\vec{s}\mid\vec{t})$ over the physical parameters 
$\vec{s} = \{\vec{s}_\mathrm{orb},\,\vec{s}_\mathrm{theory}\}$, where 
$\vec{s}_\mathrm{orb} = \{m_p,\, m_c,\, P_b,\, e,\, \omega,\, \sin i,\, d,\, w_p,\, w_q,\, w_h\}$ 
collects the binary and geometrical parameters (component masses, orbital period, 
eccentricity, periastron argument, inclination, distance, and peculiar velocity 
components), and $\vec{s}_\mathrm{theory} = \{\alpha_1,\, \alpha_2,\, c_\omega,\, s_1, s_2, \sigma_1', \sigma_2'\}$ 
collects the Einstein--\ae ther coupling constants and the pulsar sensitivity 
parameters.

In Section~\ref{sec:post-Keplerian} we showed how post-Keplerian effects
can be evaluated in Einstein-\ae ther up to 1PN order and reported explicit
expressions. Furthermore, we identify the preferred frame singled out by the \ae ther field with the frame 
in which the cosmic microwave background (CMB) is isotropic. The components of the peculiar velocity $\bm{w}$ of the system in the CMB rest frame are  connected to $\{\mu_\alpha,\mu_\delta, v_r\}$ through the transformation discussed in Appendix~\ref{app:vel_transf}. 
Those expressions, combined with standard identities relating
orbital elements, define a forward map
\begin{equation}
    \vec{\theta}(\vec{s}) = \bigl(\,\vec{\theta}_\mathrm{orb}(\vec{s}_\mathrm{orb}),\;
    \vec{\theta}_\mathrm{PK}(\vec{s})\,\bigr),
\end{equation}
where $\vec{\theta}_\mathrm{orb}(\vec{s}_\mathrm{orb})$ is an algebraic
re-parameterisation and $\vec{\theta}_\mathrm{PK}(\vec{s})$ encodes the
full 1PN Einstein-\ae ther predictions. If the inverse function $\vec{s}(\vec{\theta})$ were known, a simple approach would be to invert the map $\vec{\theta}(\vec{s})$ and
push forward the samples from $p(\vec{\theta}\mid\vec{t})$ obtained with \textsc{Vela}. Unfortunately, $\vec{s}$ has more components than
$\vec{\theta}$, so the forward map $\vec{\theta}(\vec{s})$ is many-to-one
and no unique inverse exists. The correct procedure in this case follows from a change of variables in the posterior.
Given that we know $\vec{\theta}(\vec{s})$ analytically from the
post-Keplerian relations derived in Sec.~\ref{sec:post-Keplerian}, we can
write
\begin{equation}
    p(\vec{s}\mid\vec{t})
    \propto
    p\!\left(\vec{\theta}(\vec{s})\mid\vec{t}\right)
    \frac{\pi(\vec{s})}{\pi\!\left(\vec{\theta}(\vec{s})\right)},
\label{eq:resampling}
\end{equation}
where $\pi(\vec{s})$ and $\pi(\vec{\theta})$ are the priors on the theory
and timing parameters, respectively. Equation~\eqref{eq:resampling} is
the standard importance-reweighting formula used to change priors on an
existing posterior~\cite{Smith1992BayesianSW}; here we additionally perform
a change of parameterisation, mapping the posterior from $\vec{\theta}$-
to $\vec{s}$-space through the analytically known forward model.

The key practical challenge of Eq.~\eqref{eq:resampling} is that
$p(\vec{\theta}(\vec{s})\mid\vec{t})$
must be evaluated at arbitrary points $\vec{s}$, whereas \texttt{Vela} provides only a finite set of weighted samples from $p(\vec{\theta}\mid\vec{t})$. We therefore fit a
continuous density estimator to the timing posterior samples before
performing the resampling. Specifically, we train a \emph{normalizing
flow}~\cite{Kobyzev:2019ydm} on the \texttt{Vela} output. A normalizing flow
is a bijective, differentiable transformation that maps a simple base
distribution (here a standard Gaussian) to an arbitrarily complex target,
and whose log-density can be evaluated exactly at any point in parameter
space. Once trained, the flow provides a smooth, tractable approximation
$\hat{p}(\vec{\theta}\mid\vec{t})$ that can be queried at any
$\vec{\theta}(\vec{s})$ required by the resampling integral. In practice, we draw samples $\vec{s}_i$ from the prior $\pi(\vec{s})$,
evaluate the unnormalised weight
\begin{equation}
    w_i =
    \frac{\hat{p}\!\left(\vec{\theta}(s_i)\mid\vec{t}\right)}
         {\pi\!\left(\vec{\theta}(s_i)\right)},
\label{eq:weights}
\end{equation}
and treat the weighted set $\{s_i, w_i\}$ as a Monte Carlo
representation of $p(\vec{s}\mid\vec{t})$.
For the normalizing flow we use the implementation provided by
\texttt{floZ}~\cite{Srinivasan:2024uax}, a robust
Bayesian evidence
estimator.

\section{Results and constraints}
\label{sec:results}

Here we present the main results from the timing analysis and the projected
constraints on Einstein-\ae ther. Our timing solution for PSR~J1738+0333, 
obtained via 
The result of the Bayesian inference on the ToAs is presented in 
Table~\ref{tab:timing-results}, which reports the median and half the 68\% credible 
interval of the marginalized posteriors for the timing parameters, while the full 
posteriors for the orbital parameters are shown in Figure~\ref{fig:corner_post_keplerian}.

The analysis proceeded in two stages. First, we performed a preliminary 
linear fit of the ToAs with \textsc{pint} against a timing model including 
spin, astrometric, and Keplerian and post-Keplerian orbital parameters. 
A cross-check with {\tt TEMPO2}~\citep{Hobbs:2006zg} yielded consistent results. 
The best-fit values and uncertainties from this linear fit were then used 
to initialise a full Bayesian non-linear fit with \textsc{Vela}, which 
simultaneously models the timing solution together with white noise 
(\texttt{EFAC}, \texttt{EQUAD}, \texttt{ECORR}), and power-law models 
for red noise and stochastic DM variations as described in Sec.~\ref{sec:bayesian_timing}. 

For the timing parameters, we adopted Gaussian priors centred on the 
linear-fit values with standard deviations scaled by a factor $\xi$ 
with respect to the linear-fit uncertainties. This is necessary because 
the posterior support for well-measured timing parameters is extremely 
narrow compared to the prior volume, making agnostic priors highly 
inefficient for sampling. We set $\xi = 100$ for all timing parameters 
except \texttt{DM}, \texttt{DM1} and \texttt{DM2} (the constant deterministic DM and its first and second time derivatives), for which we used $\xi = 10^{4}$; the larger value 
accounts for the fact that the inclusion of a stochastic component 
--- absent in the preliminary linear fit --- can shift the mean \texttt{DM} 
value appreciably. All noise and nuisance parameters (\texttt{EFAC}, 
\texttt{EQUAD}, \texttt{ECORR}, \texttt{TNREDAMP}, \texttt{TNREDGAM}, 
\texttt{TNDMAMP}, \texttt{TNDMGAM}, \texttt{JUMP}, \texttt{PHOFF}) 
were assigned the default priors implemented in \textsc{Vela}~\cite{Susobhanan:2024yef}. For both red noise and DM, we fixed the number of harmonics to $N_f=44$, consistently with the IPTA analysis~\cite{Antoniadis:2022pcn}. We have checked that such a high number is actually required. 

The timing solution presented in Table~\ref{tab:timing-results} is in good 
agreement with the previous analysis of~\cite{Freire:2012}. The measured 
orbital period derivative, $\dot{P}_b = (-1.82 \pm 0.25) \times 10^{-14}$,
represents an improvement over the uncertainty reported
in~\cite{Freire:2012}. We note that the linear fit yields an even tighter 
constraint, $\dot{P}_b = (-1.61 \pm 0.14) \times 10^{-14}$ (improving over Ref.~\cite{Freire:2012} by a factor of $\sim2$); however, 
the Bayesian analysis, by fully accounting for parameter correlations, 
produces a more conservative and statistically robust result, which we 
adopt for the subsequent theory constraints. To derive the intrinsic orbital period derivative, $\dot{P}^{\mathrm{corr}}_b$, we subtract the Galactic and Shklovskii contributions from Eq.~\eqref{eq:pbdot_kin}, adopting the Galactic model described in Appendix~\ref{app:pbdot_gal}.
Using the median values of the posterior distributions, we obtain $\dot{P}_b^{\mathrm{Gal}} = (-3.0 \pm 0.3)\times 10^{-16}$ and $\dot{P}_b^{\mathrm{Shkl}} = (9.3 \pm 0.6)\times 10^{-15}$. The kinematic correction is thus dominated by the Shklovskii term.

\begin{table}[htbp]
\centering
\begin{tabular}{lcc}
\hline\hline
Parameter & Prior & Range / $(\mu,\,\sigma)$ \\
\hline
\multicolumn{3}{l}{\textit{Orbital parameters}} \\
$P_b$ (d)            & Uniform   & $[0.3540,\; 0.3553]$ \\
$x$ (lt-s)           & Uniform   & $[0.3434,\; 0.3435]$ \\
$\eta_{LL}$         & Uniform   & $[-10^{-6},\; 10^{-6}]$ \\
$\kappa_{LL}$         & Uniform   & $[-10^{-6},\; 10^{-6}]$ \\
$\cos i$             & Uniform   & $[0.5,\; 0.9]$ \\
$\mu_\alpha$ (mas/yr)& Uniform   & $[5,\; 9]$ \\
$\mu_\delta$ (mas/yr)& Uniform   & $[3,\; 8]$ \\
$d$ (kpc)            & Uniform  & $[1,\; 2]$ \\
\hline
\multicolumn{3}{l}{\textit{Einstein-\ae ther parameters}} \\
$\alpha_1$     & Uniform   & $[-2\times10^{-4},\; 0]$ \\
$\alpha_2$     & Uniform   & $[-10^{-6},\; 0]$ \\
$c_\omega$           & Uniform   & $[0,\; 10^{-3}]$ \\
$\sigma_1'$               & Uniform   & $[0,\; 10^{-4}]$ \\
\hline
\multicolumn{3}{l}{\textit{Optical parameters}} \\
$m_c\;(M_\odot)$     & Gaussian  & $(0.181,\; 0.008)$ \\
$v_r$ (km/s)         & Gaussian  & $(-42,\; 16)$ \\
\hline\hline
\end{tabular}
\caption{Prior distributions adopted for the theory inference.
Uniform priors are specified by their support $[a,\, b]$;
Gaussian priors by their mean and standard deviation $(\mu,\,\sigma)$.}
\label{tab:priors_theory}
\end{table}

We now turn to the constraints on the Einstein-\ae ther parameters.
We sampled from the posterior in Eq.~\eqref{eq:resampling} using \texttt{emcee}~\cite{Foreman-Mackey:2012any} with $10\times n_\mathrm{dim}$ walkers
and $n_\mathrm{step} = 4\times10^4$ steps per walker. The model contains eighteen parameters in principle: the distance, the two bare masses
$\tilde{m}_{p},\tilde{m}_{c}$, the orbital period $P_b$ and eccentricity $e$, the
three orbital angles $\Omega$, $\iota$, and $\omega$, the sensitivities
and their first derivatives $s_{1,2}$ and $s_{1,2}'$, the three
Einstein-\ae ther coupling constants $\alpha_1$, $\alpha_2$,
and $c_\omega$, and the three components of the centre-of-mass velocity
$\{w_p,w_q,w_h\}$. However, the pulsar sensitivity $s_1$ can be inferred from the pulsar mass via the 
fit of Eq.~(80) in~\cite{Gupta_2021}, which expresses $s_1$ as a function of 
the compactness. Since the NS radius depends only weakly on the equation 
of state (EoS) for the measured value of $m_p$, the choice of EoS has a negligible 
impact on our results; we therefore adopt the \texttt{APR} EoS~\cite{Akmal:1998cf} 
for definiteness. To quantify the residual EoS dependence, we compute the relative difference in the predicted $\dot{P}_b$ with respect to the APR value, using our best estimates of $m_p$, $m_c$, $P_b$, and $e$, across the full parameter space of $\alpha_1$, $\alpha_2$, and $c_\omega$ within their prior ranges. For softer EoS models, such as WFF1~\cite{PhysRevC.38.1010} and SLY4~\cite{Douchin:2001sv}, this difference remains below 1.5\% throughout.

\begin{figure}[htbp]
    \centering
    \includegraphics[width=0.9\columnwidth]{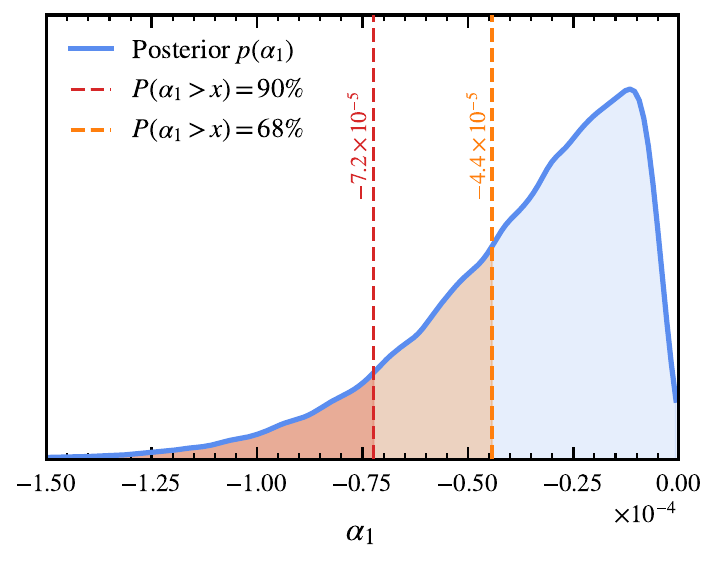}
    \caption{
        Posterior probability density of $\alpha_1$ for PSR J1738$+$0333. The shaded regions and dashed vertical lines mark the one-sided lower bounds at $68\%$ (orange) and $90\%$ (red) credible level.}
    \label{fig:alpha1_bound}
\end{figure}

Two simplifications further reduce the number of
sampled parameters to fourteen. First, since the companion is a WD, its sensitivity and sensitivity derivative are negligible and we
set $s_2 = \sigma_2' = 0$. Second, the longitude of the ascending node
$\Omega$ is not directly accessible from pulsar timing and is expected
to be weakly correlated with the remaining parameters; we therefore fix
$\Omega = 0$. We have verified that varying this choice does not
appreciably affect our results. To reduce the correlation between the sampled parameters we have sampled the set $\{\mu_\alpha,\mu_\delta, v_r\}$ instead of directly $\{w_p,w_q,w_h\}$, using the map in Appendix~\ref{app:vel_transf}. 

The priors adopted for all free
parameters are listed in Table~\ref{tab:priors_theory}.
Finally, by fitting the periodic radial velocity curve of the WD companion 
(Fig.~2 of Ref.~\cite{Antoniadis:2012vy}), one can simultaneously determine the 
radial-velocity amplitude $K_\mathrm{WD}=171 \pm 5$\,km\,$s^{-1}$ and the systemic radial velocity $v_r$ 
of the binary centre of mass. Combined with the pulsar's orbital semi-amplitude 
$K_\mathrm{PSR} = 2\pi a_1 \sin i / (P_b \sqrt{1-e^2})$, this yields the mass ratio 
$q = K_\mathrm{WD}/K_\mathrm{PSR}$~\citep{Antoniadis:2012vy}, 
which we implement as a Gaussian constraint in our MCMC analysis.

The complete marginalised posterior distributions over the theory
parameters are shown in Figure~\ref{fig:corner_theory}. Most of them
are only weakly constrained by the binary timing data, with their 
posteriors closely tracking the prior. This is expected: 
$\alpha_2$ is already tightly bounded by Solar System 
experiments~\cite{Will:2014kxa,muller:2005sr}, so its prior is sufficiently narrow that 
the binary timing data add little further information. The coupling 
$c_\omega$ enters the two-body dynamics only weakly through 
sensitivity-dependent terms, and the present dataset does not yield 
meaningful constraints beyond existing bounds; the same applies to $\sigma_1'$. 

The exception is $\alpha_1$, whose posterior is clearly pulled away from the 
prior and concentrates near the general-relativistic value $\alpha_1 = 0$, 
indicating that the data carry genuine constraining power on this parameter. 
Since $\alpha_1 = 0$ lies at the boundary of the prior, only a one-sided bound 
is meaningful, and we therefore quote one-sided credible intervals, also shown in Figure~\ref{fig:alpha1_bound}, which displays the full marginalized posterior:
\begin{equation}
\begin{split}
    \alpha_1 > -4.4 \times 10^{-5} \quad (68\%), \\
    \alpha_1 > -7.2 \times 10^{-5} \quad (90\%).
    \end{split}
    \label{eq:alpha1_bound}
\end{equation}
We note that the $68\%$ bound is stronger than the result of~\citet{Gupta_2021}, 
although of comparable magnitude. Furthermore, the increased completeness of the 
present analysis partially offsets the improvement from the more precise measurement 
of $\dot{P}_b$: in particular, we accounted for the contribution of the 
centre-of-mass peculiar velocity $\mathbf{w}$ to $\dot{P}_b$, which was neglected 
(set to zero) in previous works, in tension with the proper motion and radial 
velocity measurements. The net result is a constraint on a firmer statistical 
footing.

\section{Conclusions}
\label{sec:conclusions}

We have presented a comprehensive timing and gravity-test analysis of the 
pulsar--WD binary PSR~J1738$+$0333, combining an extensive ToA 
dataset with a fully Bayesian inference pipeline.

The timing analysis yields a precise set of system parameters, 
summarised in Table~\ref{tab:timing-results}. The Bayesian posterior on 
$\dot{P}_b$ is broader than the formal uncertainty of the preliminary 
linear fit, a feature that is expected and desirable: by marginalising 
over the full noise model and propagating parameter correlations, it 
produces a statistically more reliable interval. Besides using a full Bayesian approach, a key methodological 
advance with respect to previous analyses is the treatment of the 
peculiar velocity of the pulsar center of mass: rather than setting it to zero, 
we treated the peculiar velocity as a derived quantity constrained by the 
measured proper motion and line-of-sight velocity $v_r$, removing a 
previously unquantified systematic from the inference chain.

On the gravity-test side, $\alpha_2$ and $c_\omega$ are not 
meaningfully constrained by the binary timing data alone, as expected 
given existing Solar System bounds and the weak coupling of $c_\omega$ 
to the two-body dynamics. For $\alpha_1$, we obtain the bounds 
quoted in Eq.~\eqref{eq:alpha1_bound}, which we argue represent the most 
robust constraint derived from a single binary pulsar system to date. 
The improvement over~\citet{Gupta_2021} stems not from the dataset alone, 
but from the methodology: the full Bayesian resampling in theory-parameter 
space accounts for correlations among all post-Keplerian observables and 
correctly propagates the peculiar velocity uncertainty, both of which 
were neglected in previous work.

The framework developed here is general and directly applicable to other binary pulsar systems and alternative gravity theories, opening the way to systematic, correlation-aware tests of gravity with the growing catalog of relativistic binaries. In particular, the most stringent future constraints on Einstein--\ae ther theory are expected from a coordinated analysis of the most relativistic pulsar--WD systems, such as PSR J0348+0432~\cite{Lynch+J0348-2013} and PSR J1141-6545~\cite{Kaspi+J1141-2000}, in addition to PSR J1738+0333 analyzed here. The Double Pulsar PSR J0737-3039A/B~\cite{Burgay+J0737-2003,Lyne+J0737-2004}, despite comprising two NSs of similar mass, offers exceptional timing precision and could provide competitive bounds~\cite{Kramer+J0737-2021}.

\begin{acknowledgments}
The authors are grateful to Francesco Iraci for valuable discussions on DM noise characterization, to Joris Verbiest for helpful feedback on the timing results, which improved the analysis, and to Rahul Srinivasan for assistance with the normalizing flow implementation.

We acknowledge support from the PRIN 2022 grant ``GUVIRP - Gravity tests in the UltraViolet and InfraRed with Pulsar timing'', from the European Union’s Horizon ERC Synergy Grant ``Making Sense of the Unexpected in the Gravitational-Wave Sky'' (Grant No. GWSky-101167314, to E.B. and M.V.), and the EU Horizon 2020 Research and Innovation Programme under the Marie Sklodowska-Curie Grant Agreement No. 101007855 (to E.B.).
\end{acknowledgments}

\appendix

\section{Averaging of the $\dot{P}_b$ formula over the orbital period}
\label{app:average}

The expression for $\dot{P}_b$ in the general case using Eq.~\eqref{eq:PbDot_ae} has been worked out in~\cite{Yagi_2014} and is given by
\begin{equation}
\begin{split}
&\frac{\dot P_b}{P_b} =
-3 D \left(\frac{\mathcal{G} G a \mu m}{r_{12}^{4}}\right)
\Bigg\{
\frac{8}{15}
\left(\mathcal{A}_1 + S \mathcal{A}_2 + S^2 \mathcal{A}_3\right) 
\\
&\times
\left(12 v_{12}^2 - 11 \dot r_{12}^2\right) + 4 \left(\mathcal{B}_1 + S \mathcal{B}_2 + S^2 \mathcal{B}_3\right)\dot r_{12}^2
\\
&+ (s_1 - s_2)^2
\Bigg[
\mathcal{C}
+ \left(\frac{18}{5} \mathcal{A}_3 + 2D\right) V_{\rm CM}^j V_{\rm CM}^j
\\
&+ \left(\frac{6}{5} \mathcal{A}_3 + 36 \mathcal{B}_3 - 2\mathcal{D}\right)
\left(V_{\rm CM}^i \hat n_{12}^i\right)^2
\Bigg]
\\
&+ (s_1 - s_2)
\Bigg[
12(\mathcal{B}_2 + 2 S \mathcal{B}_3)
V_{\rm CM}^i \hat n_{12}^i
\, v_{12}^j \hat n_{12}^j
\\
&+ \frac{8}{5}(\mathcal{A}_2 + 2 S \mathcal{A}_3)
V_{\rm CM}^i
\left(3 v_{12}^i
- 2 \hat n_{12}^i v_{12}^j \hat n_{12}^j\right)
\Bigg]
\Bigg\}.
\label{eq:pbdot}
\end{split}
\end{equation}
The computation of the pulsar orbital decay however involves a systematic averaging of velocity- and
position-dependent terms in Eq.~\eqref{eq:pbdot} over one orbital period. The relative orbital
motion can be described using the semi-major axis $a$, eccentricity $e$,
and eccentric anomaly $E$, with the standard relations
\begin{align}
r &= a (1 - e \cos E), \\
\dot r &=
\frac{\sqrt{\mathcal{G} M / a}\; e \sin E}{1 - e \cos E}, \\
v^2 &= \mathcal{G} M
\left(\frac{2}{r} - \frac{1}{a}\right).
\end{align}

The position vector in the orbital plane is
\begin{equation}
\mathbf{x} =
-a
\begin{pmatrix}
\cos E - e \\
\sqrt{1-e^2}\,\sin E \\
0
\end{pmatrix},
\qquad
\mathbf{n} = \frac{\mathbf{x}}{r}.
\end{equation}

Notice that vectors are written in the peri-focal frame, and at the
periastron ($E=0$) the position vector has only the $x$ component.
The eccentric anomaly evolves according to
\begin{align}
\dot E &= 
\frac{\sqrt{\mathcal{G} M / a^3}}
     {1 - e \cos E}, \\
\tau(E) &=
\int_0^E \frac{dE'}{\dot E}
= \frac{E - e \sin E}
       {\sqrt{\mathcal{G} M / a^3}} .
\end{align}

The orbital period is therefore
\begin{equation}
P = \tau(2\pi)
  = \frac{2\pi}{\sqrt{\mathcal{G} M / a^3}} .
\end{equation}

To compute orbital averages of functions depending on the relative
velocity $\mathbf{v}_{12}$, separation $\mathbf{r}_{12}$, and
center-of-mass velocity $\mathbf{V}_{\rm CM}$, we write
\begin{equation}
\label{eq:average}
\langle f \rangle =
\frac{1}{P}\int_0^P f\,dt
=
\frac{1}{P}\int_0^{2\pi}
\frac{f}{\dot E}\,dE .
\end{equation}

Defining $\mathbf{V}_{\rm CM} = (w_p,w_q,w_h)$ and evaluating the
averages using Eq.~\eqref{eq:average} yields for the different terms
\begin{align}
&\left\langle\frac{v_{12}^2}{r_{12}^4}\right\rangle
=
\frac{(2+7e^2+e^4)\,\mathcal{G}M}
     {2a^5(1-e^2)^{7/2}}, \\[10pt]
&\left\langle\frac{\dot r_{12}^2}{r_{12}^4}\right\rangle
=
\frac{e^2(4+e^2)\,\mathcal{G}M}
     {8a^5(1-e^2)^{7/2}}, \\[10pt]
&\left\langle\frac{\mathbf{V}_{\rm CM}\!\cdot\!
      \mathbf{V}_{\rm CM}}{r_{12}^4}\right\rangle
=
\frac{(2+e^2)(w_p^2+w_q^2+w_h^2)}
     {2a^4(1-e^2)^{5/2}}, \\[10pt]
&\left\langle\frac{(\mathbf{V}_{\rm CM}\!\cdot\!
       \mathbf{n}_{12})^2}{r_{12}^4}\right\rangle
=
\frac{(4+3e^2)w_p^2 + (4+e^2)w_q^2}
     {8a^{4}(1-e^2)^{5/2}}, \\[10pt]
&\left\langle\frac{(\mathbf{v}_{12}\!\cdot\!\mathbf{n}_{12})
      (\mathbf{V}_{\rm CM}\!\cdot\!\mathbf{n}_{12})}
     {r_{12}^4}\right\rangle
=
\frac{e(4+e^2)\sqrt{\mathcal{G}M}\,w_q}
     {8a^{9/2}(1-e^2)^3}, \\[10pt]
&\left\langle\frac{\mathbf{V}_{\rm CM}\!\cdot\!\mathbf{v}_{12}}
     {r_{12}^4}\right\rangle
=
\frac{e(4+e^2)\sqrt{\mathcal{G}M}\,w_q}
     {2a^{9/2}(1-e^2)^3}.
\end{align}

The average of the term $C_a/r_{12}^4$ is more involved: since it is a 0PN contribution, the average must be performed on the 1PN-corrected orbit. The leading $1/r^4$ term is expanded to first
PN order as
\begin{align}
r_{\rm 1PN}
&= a(E)\,[1 - e(E)\cos E], \\
a(E)
&= a_0 +
\frac{\Delta a}{2\pi}
\frac{E}{c^2}, \\
e(E)
&= e_0 +
\frac{\Delta e}{2\pi}
\frac{E}{c^2}.
\end{align}

The expansion of the inverse separation becomes
\begin{align}
\frac{1}{r_{\rm 1PN}^4}
\simeq
&\frac{1}
 {a_0^4(1-e_0\cos E)^4}
\nonumber \\
&-
\frac{2E(-\Delta a + e_0\Delta a\cos E
+ a_0\Delta e\cos E)}
     {a_0^5 c^2 \pi (1-e_0\cos E)^5}.
\end{align}

The time-averaged PN correction is then written as
\begin{equation}
\left\langle\frac{C_a}{r_{12}^4}\right\rangle
=
C_a (I_a + I_b),
\label{eq:leading_average}
\end{equation}
where we called $I_a$ the Newtonian contribution and
$I_b$ the PN correction. The semi-major axis $a$ is related to the observed
orbital period $P_b$ through
\begin{equation}
P_b =
\frac{2\pi}{\sqrt{\mathcal{G}M/a^3}}
\quad \Rightarrow \quad
a =
\left(
\frac{\mathcal{G}M P_b^2}{4\pi^2}
\right)^{1/3}.
\end{equation}
Replacing $a$ in favor of $P_b$, the terms in Eq.~\eqref{eq:leading_average} become
\begin{align}
&I_a=\frac{2^{5/3}\pi^{8/3}(2+e^2)}
 {(1-e^2)^{5/2}M^{4/3}P_b^{8/3}}
\nonumber\\
&I_b= \frac{\pi^{10/3}}{8\cdot 2^{2/3}\,M^{5/3}\,P_b^{10/3}}
\left(\mathcal{P}(e)\,\delta a_{\rm sec} + \mathcal{Q}(e)\,\delta e\right)
\label{eq:averages}
\end{align}
where $\mathcal{P}(e)$ and $\mathcal{Q}(e)$ are defined in Sec.~\ref{sec:post-Keplerian}.

Applying the same substitution to all the terms in Eqs.~\eqref{eq:averages} and combining them we get to the final expression for the eccentric $\dot{P}_b$, which is given by Eq.~\eqref{eq:pbdot_avg}. This accounts for the effect of sensitivities and PN contributions, as well as
effects of the pulsar center-of-mass motion, on orbital decay.

\section{Galactic contribution to $\dot{P}_b$}
\label{app:pbdot_gal}

To accurately correct the observed orbital period derivative $\dot{P}_b$ 
for Galactic acceleration, we computed the line-of-sight acceleration 
using the \texttt{galpy} package with the 
\texttt{MWPotential2014} Milky Way potential~\cite{Bovy:2014vfa}, which provides a 
self-consistent three-dimensional model including contributions from the 
Galactic bulge, disk, and dark matter halo.

The pulsar's equatorial coordinates $(\alpha, \delta)$, determined from 
the timing solution, are first converted to Galactic coordinates $(l, b)$ 
using \texttt{Astropy}~\cite{2022ApJ...935..167A}. Together with the 
distance $d$, these are used to compute the Galactocentric cylindrical 
radius
\begin{equation}
    R = \sqrt{R_0^2 + (d\cos b)^2 - 2R_0\, d\cos b\cos l}
    \label{eq:galR}
\end{equation}
and the vertical height $z = d \sin b$ above the Galactic plane, where 
$R_0$ is the Sun--Galactic centre distance. We adopt 
$R_0 = 8.2\,\mathrm{kpc}$ and a local circular velocity $\Theta_0 = 
230\,\mathrm{km\,s^{-1}}$~\cite{10.1093/mnras/stac3529,2023A&A...676A.134P}, to scale the potential and 
convert the dimensionless \texttt{galpy} forces to physical units.

The radial and vertical components of the Galactic acceleration, $a_R$ 
and $a_z$, are then projected along the line of sight as
\begin{align}
    a_\mathrm{Gal} = &a_R \frac{d\cos^2 b - R_0 \cos b \cos l}{R} \nonumber \\
    &\qquad \qquad + a_{R,\odot}\cos b \cos l + a_z \sin b,
    \label{eq:agal_proj}
\end{align}
where $a_{R,\odot}$ is the radial Galactic acceleration at the Sun's 
location. The Galactic contribution to the period derivative then follows 
from Eq.~\eqref{eq:pbdot_kin} with $a_\mathrm{Gal}$ as defined above.

\begin{figure}[htbp]
    \centering
    \includegraphics[width=\columnwidth]{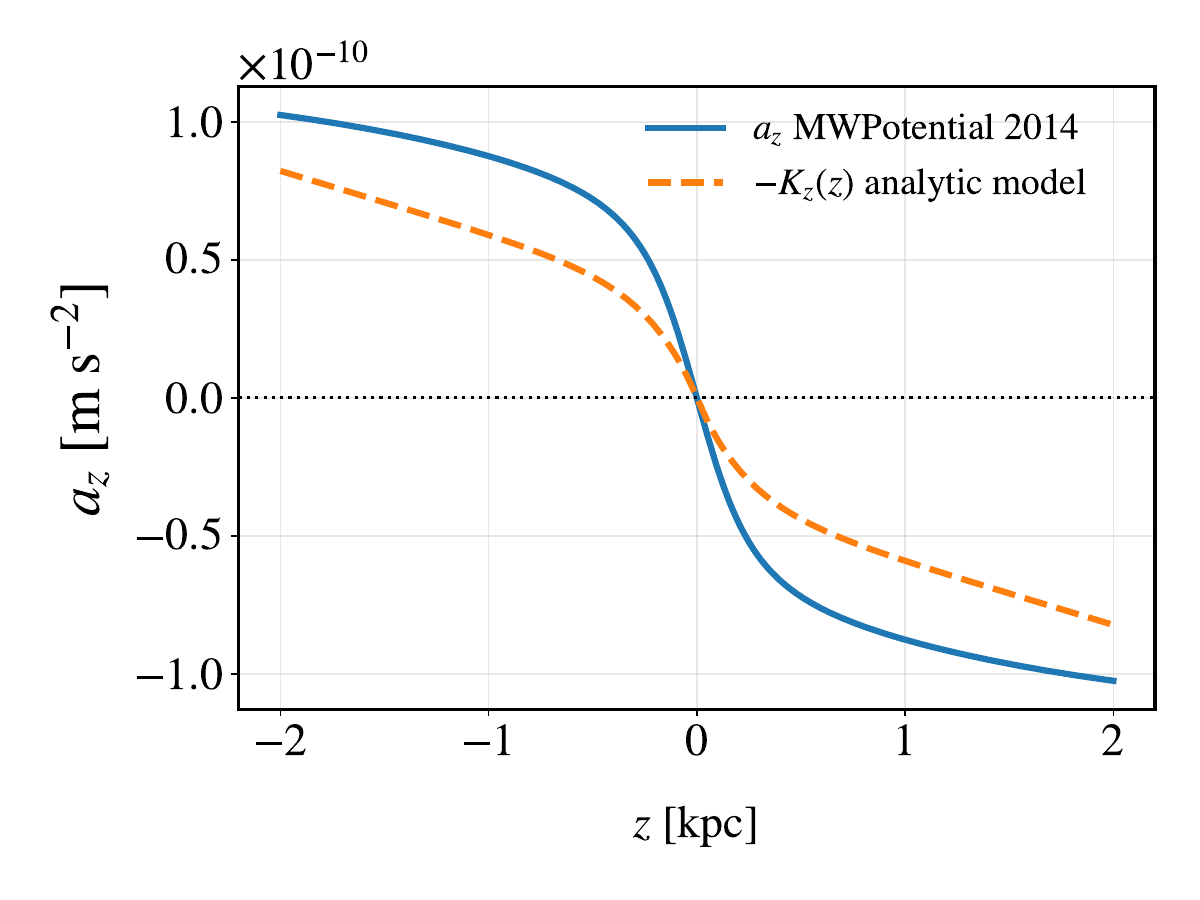}
    \caption{Comparison between the Galactic acceleration along the line 
    of sight computed with \texttt{MWPotential2014} via \texttt{galpy} 
    and the analytic $K_z$ model of~\citet{2009MNRAS.400..805L}, as a 
    function of distance $d$ to the pulsar. The \texttt{galpy}-based 
    estimate is systematically larger in amplitude, reversing the sign 
    of $\dot{P}_b^\mathrm{Gal}$ with respect to~\citet{Freire:2012}.} 
    \label{fig:galactic_potential}
\end{figure}

For comparison, we also evaluated the simpler analytic model of~\citet{2009MNRAS.400..805L} for the vertical acceleration,
\begin{equation}
    a_\mathrm{z,\,analytic} = -\frac{K_z(z)}{c}\,|\sin b|,
    \label{eq:kz}
\end{equation}
where $K_z(z)$ parameterises the vertical force near the Galactic plane. 
This model captures the overall vertical structure but neglects radial 
variations and the detailed Galactic potential. As shown in 
Figure~\ref{fig:galactic_potential}, the \texttt{MWPotential2014} result 
exceeds the analytic $K_z$ estimate in amplitude, which reverses the sign 
of $\dot{P}_b^\mathrm{Gal}$ compared to~\citet{Freire:2012}. Nevertheless, 
as discussed in Section~\ref{sec:results}, the Shklovskii term dominates 
the kinematic correction for this system.

\section{Complete corner plots}
\label{app:corner}
\begin{figure*}[htbp]
    \centering
    \includegraphics[width=\textwidth]{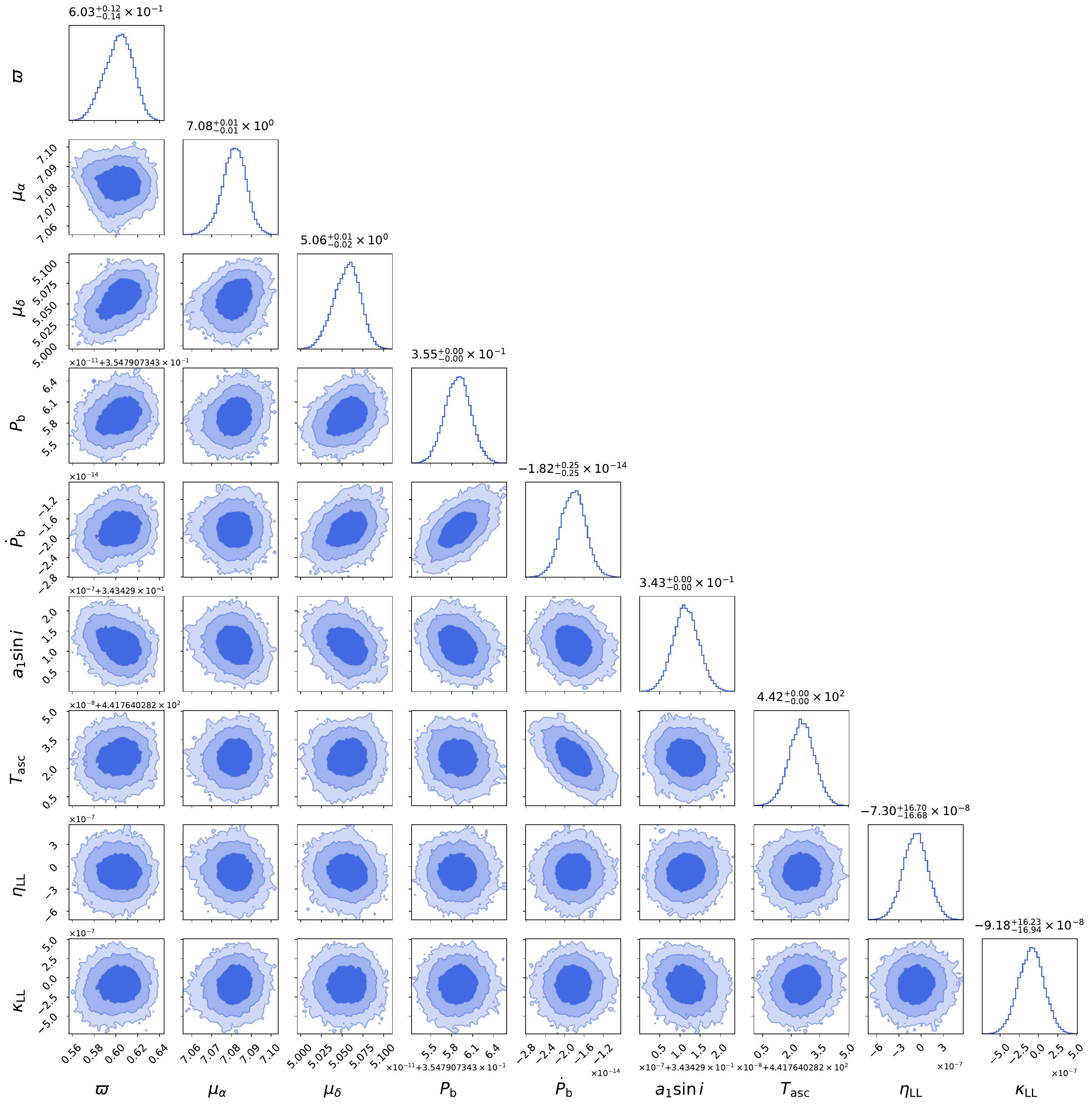}
    \caption{
        Corner plot of the joint posterior distributions for the post-Keplerian parameters.
        The off-diagonal panels show the marginalized two-dimensional posteriors, with
        68\%, 95\%, and 99.7\% credible intervals indicated.
        The diagonal panels display the corresponding one-dimensional projections.
    }
    \label{fig:corner_post_keplerian}
\end{figure*}

\begin{figure*}[htbp]
    \centering
    \includegraphics[width=\textwidth]{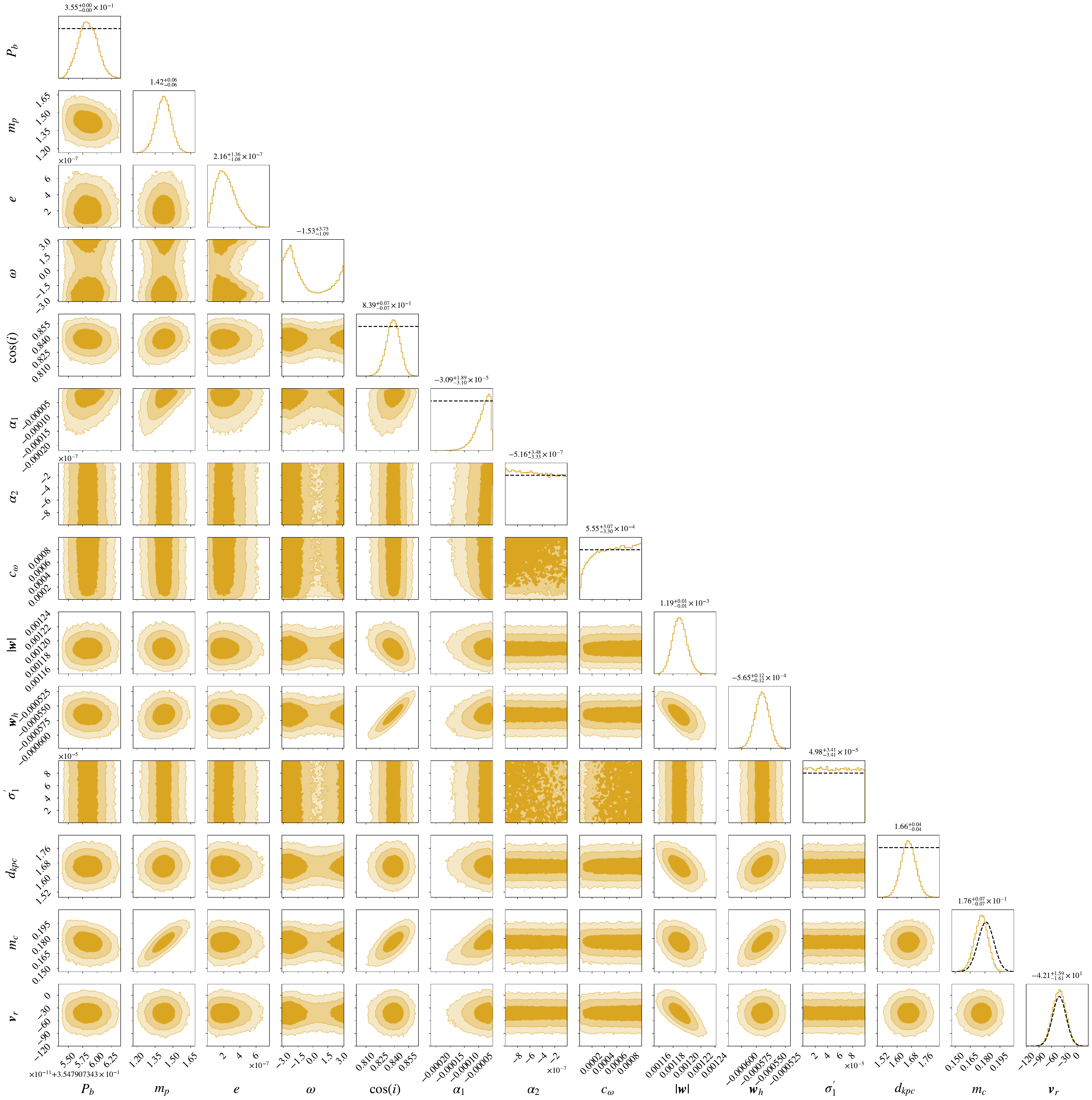}
    \caption{
        Corner plot of the joint posterior distributions for the theory parameters. The off-diagonal panels show the marginalized two-dimensional posteriors, with
        68\%, 95\%, and 99.7\% credible intervals indicated.
        The diagonal panels display the corresponding one-dimensional projections. Black dashed lines show the chosen prior distributions for the sampled parameters 
listed in Table~\ref{tab:priors_theory}; the remaining quantities shown ($m_p$, $e$, $\omega$, 
$|\bm{w}|$, and $w_h$) are derived parameters and no prior is displayed for them. }
    \label{fig:corner_theory}
\end{figure*}

We present here the complete posterior distributions obtained at each
stage of the analysis. Figure~\ref{fig:corner_post_keplerian} shows the corner
plot of the timing posterior $P(\vec{\theta}\mid\vec{t})$, displaying
the marginalised one- and two-dimensional distributions for all orbital
and post-Keplerian parameters. Figure~\ref{fig:corner_theory} shows the
corresponding corner plot in theory-parameter space,
$P(\vec{s}\mid\vec{t})$, obtained after the resampling procedure
described in Sec.~\ref{sec:resampling}.

\section{Velocity of a binary system in the Aether frame and its projection into the orbital frame}
\label{app:vel_transf}

We briefly summarize here the relation between the orbital (perifocal) components, 
the sky-projected components, and the Cartesian velocity in the International 
Celestial Reference System (ICRS).

Let $\mathbf{w}$ be the velocity of the binary system with respect to the CMB frame,
expressed in orbital (perifocal) components $(w_p, w_q, w_h)$. The orbital frame is 
defined by the orthonormal triad $(\mathbf{e}_p, \mathbf{e}_q, \mathbf{e}_h)$, with 
$\mathbf{e}_p$ pointing toward periastron, $\mathbf{e}_q$ lying in the orbital plane, 
and $\mathbf{e}_h$ parallel to the orbital angular momentum. The corresponding 
rotation matrix to Cartesian ICRS coordinates is $O = R_3(\Omega)\,R_1(i)\,R_3(\omega)$, 
where $i$ is the inclination, $\Omega$ the longitude of the ascending node, and 
$\omega$ the argument of periastron, so that $\mathbf{w}^{(\mathrm{ICRS})} = O\,\mathbf{w}$.

In our preferred-frame analysis, $\mathbf{w}^{(\mathrm{ICRS})}$ is related to the 
velocity $\mathbf{v}^{(\mathrm{ICRS})}$ with respect to the Solar System by
\[
\mathbf{w}^{(\mathrm{ICRS})} = \mathbf{v}^{(\mathrm{ICRS})} + \mathbf{V}_{\odot\mathrm{CMB}}^{(\mathrm{ICRS})},
\]
where $\mathbf{V}_{\odot\mathrm{CMB}}$ is the velocity of the Solar System barycenter 
relative to the CMB frame. Introducing the orthonormal sky basis $(\mathbf{n}, \mathbf{e}_\alpha, \mathbf{e}_\delta)$, 
where $\mathbf{n}$ is the line-of-sight direction, $\mathbf{e}_\alpha$ points toward 
increasing right ascension, and $\mathbf{e}_\delta$ toward increasing declination, 
the sky components $(v_r, v_\alpha, v_\delta)$ are obtained by projection onto this 
basis. Since this basis is orthonormal, $S$ (whose rows are the Cartesian ICRS 
components of $\mathbf{n}$, $\mathbf{e}_\alpha$, and $\mathbf{e}_\delta$) is a pure 
rotation matrix with $S^{-1} = S^T$, so that 
$\mathbf{v}^{(\mathrm{ICRS})} = S^T(v_r, v_\alpha, v_\delta)^T$. The perifocal 
components are therefore
\[
(w_p, w_q, w_h)^T = O^T S^T (v_r, v_\alpha, v_\delta)^T + O^T \mathbf{V}_{\odot\mathrm{CMB}}^{(\mathrm{ICRS})},
\]
or equivalently, $w_p = \mathbf{e}_p \cdot (S^T(v_r, v_\alpha, v_\delta)^T + 
\mathbf{V}_{\odot\mathrm{CMB}}^{(\mathrm{ICRS})})$, and analogously for $w_q$ and $w_h$. The sky velocity components follow from the observables via $v_\alpha = K d\,\mu_\alpha$ 
and $v_\delta = K d\,\mu_\delta$, where $d$ is the distance in kpc and the conversion 
factor $K = (1\,{\rm AU})/(1\,{\rm yr}) \simeq 4.74047\,{\rm km\,s^{-1}}$ is obtained 
directly from unit conversions between mas\,yr$^{-1}$, kpc, and km\,s$^{-1}$.

\bibliographystyle{apsrev4-1}
\bibliography{refs}
\end{document}